\documentclass[a4paper]{rsproca_new_arxiv}

\DeclareGraphicsExtensions{.pdf,.jpeg,.png}
\usepackage[numbers]{natbib}
\graphicspath{}

\newcommand{\BE}{\begin{equation}}
\newcommand{\EE}{\end{equation}}
\newcommand{\BF}{\begin{figure}\centering}
\newcommand{\EF}{\end{figure}}
\newcommand{\BT}{\begin{table}\centering}
\newcommand{\ET}{\end{table}}

\newcommand{\D}[1]{\,\mathrm{d} #1} 
\newcommand{\MAT}[1]{\boldsymbol{#1}} 
\newcommand{\J}{\mathrm{i}} 

\newcommand{\RE}{\mathrm{Re}}
\newcommand{\IM}{\mathrm{Im}}

\newcommand{\WM}{W_\mathrm{m}^{\MAT{A}}}            
\newcommand{\WE}{W_\mathrm{e}^\varphi}            
\newcommand{\WMM}{W_\mathrm{m}}  
\newcommand{\WEE}{W_\mathrm{e}}  
\newcommand{\WMg}{\widetilde{W}_\mathrm{m}}   
\newcommand{\WEg}{\widetilde{W}_\mathrm{e}}   
\newcommand{\Wrad}{W_\mathrm{rad}} 
\newcommand{\Wom}{W_{\partial \omega}}      
\newcommand{\PM}{P_\mathrm{m}^{\MAT{A}}}     
\newcommand{\PE}{P_\mathrm{e}^\varphi}     
\newcommand{\PMM}{P_\mathrm{m}}    
\newcommand{\PEE}{P_\mathrm{e}}    
\newcommand{\Prad}{P_\mathrm{rad}} 
\newcommand{\Pom}{P_{\partial\omega}}      
\newcommand{\Zin}{Z_\mathrm{in}}   
\newcommand{\Pin}{P_\mathrm{in}}   

\newcommand{\QET}{\mathcal{Q}}     
\newcommand{\QEF}{Q}               
\newcommand{\QFBW}{Q_{Z}}           
\newcommand{\WETsto}{\mathcal{W}_\mathrm{sto}}
\newcommand{\WEFsto}{W_\mathrm{sto}}
\newcommand{\WEFstoT}{\widetilde{W}_\mathrm{sto}}
\newcommand{\WFBWsto}{W_\mathrm{sto}^\mathrm{FBW}}
\newcommand{\WETlost}{W_\mathrm{lost}}
\newcommand{\WEr}{W_\mathrm{r}}

\newcommand{\Ws}{W_\sigma}

\newcommand\figwidth{10}

\begin{document}
\title{Stored Electromagnetic Energy and Quality Factor of Radiating Structures}

\author{
	Miloslav~Capek$^{1}$, Lukas~Jelinek$^{1}$ and Guy~A.~E.~Vandenbosch$^{2}$}

\address{$^{1}$Department of Electromagnetic Field, Faculty of Electrical Engineering, Czech Technical University in Prague, Technicka 2, 16627, Prague, Czech Republic\\
	$^{2}$Department of Electrical Engineering, Division ESAT-TELEMIC (Telecommunications and Microwaves), Katholieke Universiteit Leuven, B-3001, Leuven, Belgium}

\subject{electromagnetism, mathematical physics, electrical engineering}

\keywords{electromagnetic theory, antenna theory, stored electromagnetic energy, quality factor}

\corres{Miloslav~Capek\\
	\email{miloslav.capek@fel.cvut.cz}}


\maketitle
\subsection*{Abstract}
	This paper deals with the old yet unsolved problem of defining and evaluating the stored electromagnetic energy -- a quantity essential for calculating the quality factor, which reflects the intrinsic bandwidth of the considered electromagnetic system. A novel paradigm is proposed to determine the stored energy in the time domain leading to the method, which exhibits positive semi-definiteness and coordinate independence, i.e. two key properties actually not met by the contemporary approaches. The proposed technique is compared with two up-to-date frequency domain methods that are extensively used in practice. All three concepts are discussed and compared on the basis of examples of varying complexity, starting with lumped RLC circuits and finishing with realistic radiators.

	\section{Introduction}
	\label{Intro}
	
	In physics, an oscillating system is traditionally characterized \cite{MorseFeshBach_MethodsOfTheoreticalPhysics} by its oscillation frequency and quality factor $Q$, which gives a measure of the lifetime of free oscillations. At its high values, the quality factor $Q$ is also inversely proportional to the intrinsic bandwidth in which the oscillating system can effectively be driven by external sources \cite{Hallen_ElectromagneticTheory,VolakisChenFujimoto_SmallAntennas}.
	

The concept of quality factor $Q$ as a single frequency estimate of relative bandwidth is most developed in the area of electric circuits \cite{Collin_FoundationsForMicrowaveEngineering} and electromagnetic radiating systems \cite{VolakisChenFujimoto_SmallAntennas}. Its evaluation commonly follows two paradigms. As far as the first one is concerned, the quality factor is evaluated from the knowledge of the frequency derivative of input impedance \cite{Rhodes_ObservableStoredEnergiesOfElectromagneticSystems,KajfezWheless_InvariantDefinitionsOfTheUnloadedQfactor,Harrington_OnTheGainAndBWofDirectAntennas}. As for the second paradigm, the quality factor is defined as $2\pi$ times the ratio between the cycle mean stored energy and the cycle mean lost energy \cite{Rhodes_ObservableStoredEnergiesOfElectromagneticSystems,IEEEStd_antennas}. Generally, these two concepts yield distinct results, but come to identical results in the case of vanishingly small losses, the reason being the Foster's reactance theorem \cite{Foster_AreactanceTheorem,Harrington_TimeHarmonicElmagField}. 

The evaluation of quality factor by means of frequency derivative of input impedance was made very popular by the work of Yaghjian and Best \cite{YaghjianBest_ImpedanceBandwidthAndQOfAntennas} and is widely used in engineering practice \cite{BestHanna_AperformanceComparisonOfFundamentalESA,Sievenpiper_ExpretimentalValidationOfPerformanceLimitsAndDesignGuidelines} thanks to its property of being directly measurable. Recently, this concept of quality factor has also been expressed as a bilinear form of source current densities \cite{CapekJelinekHazdraEichler_MeasurableQ}, which is very useful in connection with modern numerical software tools \cite{Jin_TheoryAndComputationOfElectromagneticFields}. Regardless of the mentioned advantages, the impedance concept of quality factor suffers from a serious drawback of being zero in specific circuits \cite{GustafssonNordebo_BandwidthQFactorAndResonanceModelsOfAntennas,CapekJelinekHazdra_OnTheFunctionalRelationBetweenQfactorAndFBW} and/or radiators \cite{GustafssonJonsson_AntennaQandStoredEnergiesFieldsCurrentsInputImpedance,CapekJelinekHazdra_OnTheFunctionalRelationBetweenQfactorAndFBW} with evidently non-zero energy storage. This unfortunately prevents its usage in modern optimization techniques \cite{GustafssonCismasuJonsson_PhysicalBoundsAndOptimalCurrentsOnAntennas_TAP}.

The second paradigm, in which the quality factor is evaluated via the stored energy and lost energy, is not left without difficulties either. In the case of non-dispersive components, the cycle mean lost energy does not pose a problem and may be evaluated as a sum of the cycle mean radiated energy and the cycle mean energy dissipated due to material losses \cite{Jackson_ClassicalElectrodynamics}. Unfortunately, in the case of a non-stationary electromagnetic field associated with radiators, the definition of stored (non-propagating) electric and magnetic energies presents a problem that has not yet been satisfactorily solved \cite{VolakisChenFujimoto_SmallAntennas}. The issue comes from the radiation energy, which does not decay fast enough in radial direction, and is in fact infinite in stationary state \cite{Chu_PhysicalLimitationsOfOmniDirectAntennas}. 

In order to overcome the infinite values of total energy, the evaluation of stored energy in radiating systems is commonly accompanied by the technique of extracting the divergent radiation component from the well-known total energy of the system \cite{Jackson_ClassicalElectrodynamics}. This method is somewhat analogous to the classical field \cite{Dirac_ClassicalTheoryOfRadiatingElectrons_1938} re-normalization. Most attempts in this direction have been performed in the domain of time-harmonic fields. The pioneering work in this direction is the equivalent circuit method of Chu \cite{Chu_PhysicalLimitationsOfOmniDirectAntennas}, in which the radiation and energy storage are represented by resistive and reactive components of a complex electric circuit describing each spherical mode. This method was later generalized by several works of Thal
\cite{Thal_ExactCircuitAnalysisOfSphericalWaves,Thal_QboundsForArbitrarySmallAntennas}. Although powerful, this method suffers from fundamental drawback of spherical harmonics expansion, which is unique solely in the exterior of the sphere bounding the sources. Therefore, the circuit method cannot provide any information on the radiation content of the interior region, nor on the connection of energy storage with the actual shape of radiator. 

The radiation extraction for spherical harmonics has also been performed directly at the field level. The classical work in this direction comes from Collin and Rothschild \cite{CollinRotchild_EvaluationOfAntennaQ}. Their proposal leads to good results for canonical systems \cite{CollinRotchild_EvaluationOfAntennaQ,McLean_AReExaminationOfTheFundamentalLimitsOnTheRadiationQofESA,Manteghi_FundamentalLimitsOfCylindricalAntennas}, and has been analytically shown self-consistent outside the radian-sphere \cite{Collin_MinimumQofSmallAntennas}. Similarly to the work of Chu, this procedure is limited by the use of spherical harmonics to the exterior of the circumscribing sphere.

The problem of radiation extraction around radiators of arbitrary shape has been for the first time attacked by Fante \cite{Fante_QFactorOfGeneralIdeaAntennas} and Rhodes \cite{Rhodes_AReactanceTheorem}, giving the interpretation to the Foster's theorem \cite{Harrington_TimeHarmonicElmagField} in open problems. The ingenious combination of the frequency derivative of input impedance and the frequency derivative of far-field radiation pattern led to the first general evaluation of stored energy. Fante's and Rhodes' works have been later generalized by Yaghjian and Best \cite{YaghjianBest_ImpedanceBandwidthAndQOfAntennas}, who also pointed out an unpleasant fact that this method is coordinate-dependent. A scheme for minimisation of this dependence has been developed \cite{YaghjianBest_ImpedanceBandwidthAndQOfAntennas}, but it was not until the work of Vandenbosch \cite{Vandenbosch_ReactiveEnergiesImpedanceAndQFactorOfRadiatingStructures} who, generalizing the expressions of Geyi for small radiators \cite{Geyi_AMethodForTheEvaluationOfSmallAntennaQ} and rewriting the extraction technique into bilinear forms of currents, was able to reformulate the original extraction method into a coordinate-independent scheme. A noteworthy discussion of various forms of this extraction technique can be found in the work of Gustafsson and Jonsson \cite{Gustaffson_StoredElectromagneticEnergy_PIER}. It was also Gustafsson et al. who emphasized \cite{GustafssonCismasuJonsson_PhysicalBoundsAndOptimalCurrentsOnAntennas_TAP} that under certain conditions, this extraction technique fails, giving negative values for specific current distributions. Hence the aforementioned approach remains incomplete too \cite{Vandenbosch_Reply2Comments}.

The problem of stored energy has seldom been addressed directly in the time domain. Nevertheless, there are some interesting works dealing with time-dependent energies. Shlivinski expanded the fields into spherical waves in time domain \cite{ShlivinskiHeynman_TimeQ1,ShlivinskiHeynman_TimeQ2}, introducing time domain quality factor that qualifies the radiation efficiency of pulse-fed radiators. Collarday \cite{Collardey-CalculationOfQ_FDTD} proposed a brute force method utilizing the finite differences technique. In \cite{Vandenbosch_RadiatorsInTimeDom1}, Vandenbosch derived expressions for electric and magnetic energies in time domain that however suffer from an unknown parameter called storage time. A notable work of Kaiser \cite{Kaiser_ElectromagneticInertiaReactiveEnergyAndEnergy} then introduced the concept of rest electromagnetic energy, which resembles the properties of stored energy, but is not identical to it \cite{CapekJelinek_VariousInterpretationOfTheStoredAndTheRadiatedEnergyDensity}.

The knowledge of the stored electromagnetic energy and the capability of its evaluation are also tightly connected with the question of its minimization \cite{Chu_PhysicalLimitationsOfOmniDirectAntennas,Harrington_EffectsOfAntennaSizeOnGainBWandEfficiency,Yaghjian_MinimumQforLossyAndLosslessESA,JonssonGustafsson_StoredEnergiesInElectricAndMagneticCurrentDensities_RoyA,CapekJelinekHazdraEichler_QofSphere}. Such lower bound of the stored energy would imply the upper bound to the available bandwidth, a parameter of great importance for contemporary communication devices.

In this paper, a scheme for radiation energy extraction is proposed following a novel line of reasoning in the time domain. The scheme aims to overcome the handicaps of the previously published works, and furthermore is able to work with general time-dependent source currents of arbitrary shape. It is presented together with the two most common frequency domain methods, the first being based on the time-harmonic expressions of Vandenbosch \cite{Vandenbosch_ReactiveEnergiesImpedanceAndQFactorOfRadiatingStructures} and the second using the input impedance approximation introduced by Yaghjian and Best \cite{YaghjianBest_ImpedanceBandwidthAndQOfAntennas}. All three concepts are closely investigated and compared on the basis of examples of varying complexity. The working out of all three concepts starts solely from the currents flowing on a radiator, which are usually given as a result in modern electromagnetic simulators. This raises challenging possibilities of modal analysis \cite{CapekHazdraEichler_AMethodForTheEvaluationOfRadiationQBasedOnModalApproach} and optimization \cite{Gustafsson_OptimalAntennaCurrentsForQsuperdirectivityAndRP}. 

The paper is organized as follows. Two different concepts of quality factor $Q$ that are based on electromagnetic energies (in both, the frequency and time domain), are introduced in \S\ref{EnergyConcepts}. Subsequently, the quality factor $Q$ derived from the input impedance is formulated in terms of currents on a radiator in \S\ref{FBWConcepts}. The following two sections present numerical examples: \S\ref{Sec_RLC} treats non-radiating circuits and \S\ref{Sec_Antennas} deals with radiators. The results are discussed in \S\ref{Sec_Disc} and the paper is concluded in \S\ref{Sec_Concl}.

\section{Energy concept of quality factor $Q$}
\label{EnergyConcepts}

In the context of energy, the quality factor is most commonly defined as
\BE
\label{Basic_A01}
\QEF = 2 \pi \frac{\langle\WETsto \left( t \right)\rangle}{\WETlost} = 2 \pi \frac{\WEFsto}{\WETlost},
\EE
where a time-harmonic steady state with angular frequency $\omega_0$ is assumed, with $\WETsto \left( t \right)$ as the electromagnetic stored energy, $\langle\WETsto \left( t \right) \rangle  = \WEFsto$ as the cycle mean of $\WETsto \left( t \right)$ and $\WETlost$ as the lost electromagnetic energy during one cycle \cite{Jackson_ClassicalElectrodynamics}. In conformity with the font convention introduced above, in the following text, the quantities defined in the time domain are stated in calligraphic font, while the frequency domain quantities are indicated in the roman font.

A typical $Q$-measurement scenario is depicted in figure~\ref{fig:scenario}, which shows a radiator fed by a shielded power source.
\BF
\includegraphics[width=6.5cm]{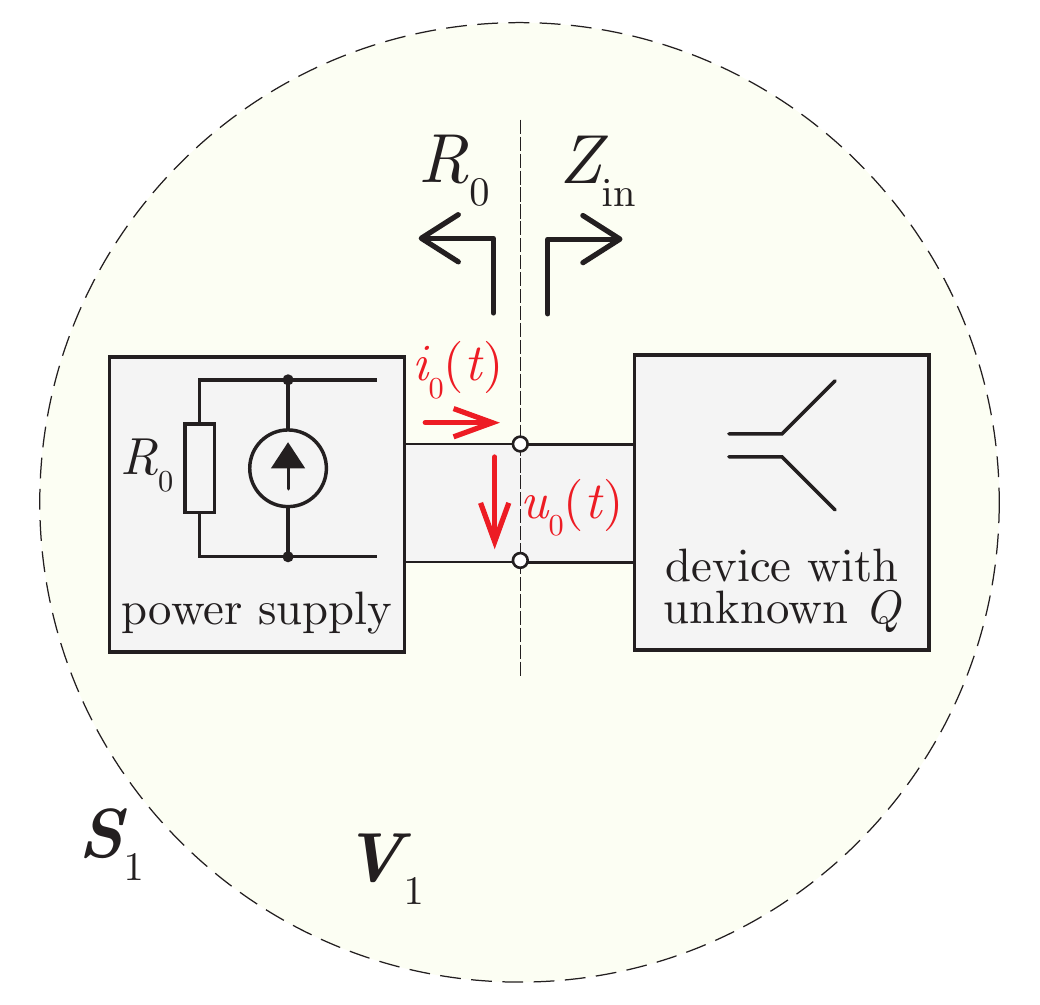}
\caption{A device with unknown $Q$ that is fed by a shielded power source with internal resistance $R_0$.}
\label{fig:scenario}
\EF
The input impedance in the time-harmonic steady state at the frequency $\omega_0$ seen by the source is $\Zin$. Assuming that the radiator is made of conductors with ideal non-dispersive conductivity $\sigma$ and lossless non-dispersive dielectrics, we can state that the lost energy during one cycle, needed for (\ref{Basic_A01}), can be evaluated as
\BE
\label{Basic_A02}
\WETlost = \int\limits_\alpha^{\alpha+T} i_0 \left( t\right) u_0 \left( t \right) \D{t} = \frac{\pi}{\omega_0} \RE\{\Zin \} I_0^2 = \WEr + \Ws,
\EE
where $I_0$ is the amplitude of $i_0 \left( t \right)$ (see figure~\ref{fig:scenario}), $\WEr$ represents the cycle mean radiation loss and $\Ws$ stands for the energy lost in one cycle via conduction. The part $\Ws$ of (\ref{Basic_A02}) can be calculated as
\BE
\label{Basic_A03}
\Ws = \frac{\pi}{\omega_0} \int\limits_{V} \sigma \left| \MAT{E} \left(\MAT{r}, \omega_0 \right) \right|^2 \D{V},
\EE
with $V$ being the shape of radiator and $\MAT{E}$ being the time-harmonic electric field intensity under the convention \mbox{$\boldsymbol{\mathcal{E}} \left(t \right) = \RE\{\MAT{E} \left( \omega \right) \mathrm{e}^{\J\omega t}\}$}, $\J = \sqrt{-1}$. At the same time, the near-field of the radiator \cite{Balanis1989} contains the stored energy $\WETsto \left( t \right)$, which is bound to the sources and does not escape from the radiator towards infinity. The evaluation of the cycle mean energy $\WEFsto$ is the goal of the following \S\ref{EnergyConcepts}\ref{TimeDomainEnergyConcept} and \S\ref{EnergyConcepts}\ref{FreqDomainEnergyConcept}, in which the power balance \cite{Harrington_TimeHarmonicElmagField} is going to be employed.

\subsection{Stored energy in time domain}
\label{TimeDomainEnergyConcept}

This subsection presents a new paradigm of stored energy evaluation. The first step consists in imagining the spherical volume $V_1$ (see figure~\ref{fig:scenario}) centred around the system, whose  radius is large enough to lie in the far-field region \cite{Balanis1989}. The total electromagnetic energy content of the sphere (it also contains heat $\Ws$) is
\BE
\label{Basic_A04}
\mathcal{W} \left( V_1, t \right) = \mathcal{W}_\mathrm{sto} \left( t \right) + \mathcal{W}_\mathrm{r} \left( V_1, t \right),
\EE
where $\mathcal{W}_\mathrm{r} \left( V_1, t \right)$ is the energy contained in the radiation fields that have already escaped from the sources. Let us assume that the power source is switched on at $t = -\infty$, bringing the system into a steady state, and then switched off at $t = t_\mathrm{off}$. For $t \in [ t_\mathrm{off},\infty)$ the system is in a transient state, during which all the energy $\mathcal{W} \left( V_1, t_\mathrm{off} \right)$ will either be transformed into heat at the resistor $R_0$ and the radiator's conductors or radiated through the bounding envelope $S_1$. Explicitly, Poynting's theorem \cite{Harrington_TimeHarmonicElmagField} states that the total electromagnetic energy at time $t_\mathrm{off}$ can be calculated as
\BE
\begin{split}
	\label{Basic_A05}
	\mathcal{W} \left( V_1, t_\mathrm{off} \right) = & \,\, R_0 \int\limits_{t_\mathrm{off}}^{\infty} i_{\mathrm{R}_0}^2 (t) \D{t} + \int\limits_{t_{\mathrm{off}}}^{\infty} \int\limits_{V} \boldsymbol{\mathcal{E}} \left(\MAT{r}, t \right) \cdot \boldsymbol{\mathcal{J}} \left(\MAT{r}, t \right) \D{V} \D{t} \\
	& + \int\limits_{t_{\mathrm{off}}}^{\infty} \oint\limits_{S_1} \Big( \boldsymbol{\mathcal{E}}_\mathrm{far} \left(\MAT{r}, t \right) \times \boldsymbol{\mathcal{H}}_\mathrm{far} \left(\MAT{r}, t \right) \Big) \cdot \D{\MAT{S}_1} \D{t},
\end{split}
\EE
in which $S_1$ lies in the far-field region.

As a special yet important example, let us assume a radiating device made exclusively of perfect electric conductors (PEC). In that case, the far-field can be expressed as \cite{Jackson_ClassicalElectrodynamics}
\begin{subequations}
	\begin{align}
		\label{Basic_A06A}
		\boldsymbol{\mathcal{H}}_\mathrm{far} \left( \MAT{r}, t \right) &= -\frac{1}{4 \pi c_0} \int\limits_{V '} \frac{\MAT{n}_0 \times \boldsymbol{\mathcal{\dot{J}}} \left( \MAT{r} ', t ' \right)}{R} \D{V '}, \\
		\label{Basic_A06B}
		\boldsymbol{\mathcal{E}}_\mathrm{far} \left( \MAT{r}, t \right) &= -\frac{\mu}{4 \pi} \int\limits_{V '} \frac{\boldsymbol{\mathcal{\dot{J}}} \left( \MAT{r} ', t ' \right) - \left( \MAT{n}_0 \cdot \boldsymbol{\mathcal{\dot{J}}} \left( \MAT{r} ', t ' \right)\right) \MAT{n}_0}{R} \D{V '}
	\end{align}
\end{subequations}
in which $c_0$ is the speed of light, $R = \left| \MAT{r} - \MAT{r} ' \right|$, \mbox{$\MAT{n}_ 0 = \left( \MAT{r} - \MAT{r} ' \right) / R$}, $t ' = t - R / c_0$ stands for the retarded time and the dot represents the derivative with respect to the time argument, i.e.
\BE
\label{Basic_A07}
\boldsymbol{\mathcal{\dot{J}}} \left( \MAT{r} ', t ' \right) = \frac{\partial \boldsymbol{\mathcal{J}} \left( \MAT{r} ', \tau \right)}{\partial \tau} \Bigg|_{\tau = t '}.
\EE
Since we consider the far-field, we can further write \cite{Balanis_Wiley_2005} $R \approx r$ for amplitudes, \mbox{$R \approx r - \MAT{r}_0 \cdot \MAT{r} '$} for time delays, with $\MAT{n}_0 \approx \MAT{r}_0$ and $r = \left|\MAT{r} \right|$. Using (\ref{Basic_A06A})--(\ref{Basic_A07}) and the above-mentioned approximations, the last term in (\ref{Basic_A05}) can be written as
\BE
\begin{split}
	\label{Basic_A08}
	\int\limits_{t_{\mathrm{off}}}^{\infty} & \oint\limits_{S_1} \Big( \boldsymbol{\mathcal{E}}_\mathrm{far} \left( \MAT{r}, t \right) \times \boldsymbol{\mathcal{H}}_\mathrm{far} \left( \MAT{r}, t \right) \Big) \cdot \MAT{r}_0 \D{S_1} \D{t} = \frac{1}{Z_0} \int\limits_{t_{\mathrm{off}}}^{\infty} \oint\limits_{S_1} \left| \boldsymbol{\mathcal{E}}_\mathrm{far} \left( \MAT{r}, t \right) \right|^2\D{S_1} \D{t} \\
	&= \frac{\mu^2}{Z_0 \left( 4 \pi \right)^2} \int\limits_{t_{\mathrm{off}}}^{\infty} \int\limits_0^\pi \int\limits_0^{2 \pi} \left| \int\limits_{V '} \Big(
	\boldsymbol{\mathcal{\dot{J}}} \left( \MAT{r} ', t ' \right) - \left( \MAT{r}_0 \cdot \boldsymbol{\mathcal{\dot{J}}} \left( \MAT{r} ', t ' \right)\right) \MAT{r}_0 \Big)
	\D{V '} \right|^2 \sin \theta \D{\varphi} \D{\theta} \D{t},
\end{split}
\EE
where $t ' = t - r / c_0 + \MAT{r}_0 \cdot \MAT{r} ' / c_0$, where $Z_0$ is the free space impedance and where the relation
\BE
\label{Basic_A06C}
\boldsymbol{\mathcal{H}}_\mathrm{far} \left( \MAT{r}, t \right) = \frac{\MAT{r}_0 \times \boldsymbol{\mathcal{E}}_\mathrm{far} \left( \MAT{r}, t \right)}{Z_0}
\EE
has been used. Utilizing (\ref{Basic_A05}) and (\ref{Basic_A08}), we are thus able to find the total electromagnetic energy inside $S_1$, see figure~\ref{fig:radExtraction_A} for graphical representation.

\BF
\includegraphics[width=13.5cm]{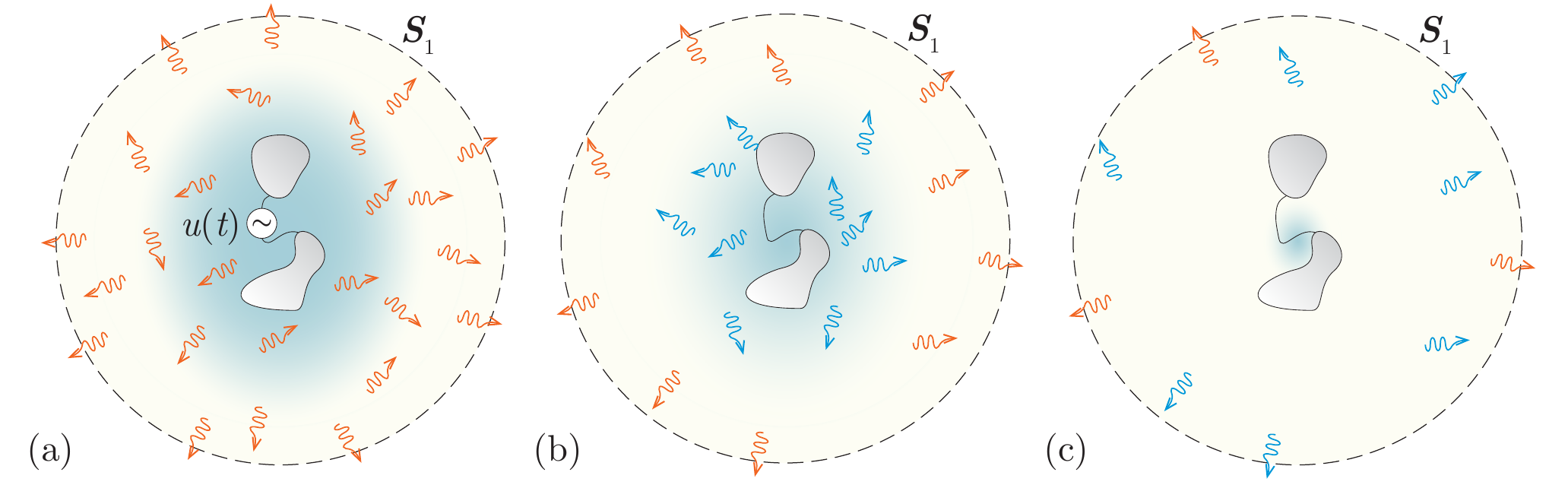}
\caption{Graphical representation of the total electromagnetic energy evaluation via (\ref{Basic_A08}) for a loss-less radiator excited by ideal voltage source. Panel (a) shows a steady state just before $t=t_{\mathrm{off}}$, when  the steady state radiation (orange wavelets) as well as the steady state stored energy (blue cloud) were maintained by the source. Panel (b) shows that after the source is switched off, the existing radiation travels to $S_1$ (and some of it also passes $S_1$) while a new radiation (blue wavelets) emerges at the expense of the stored energy. Panel (c) depicts the time $t \gg t_{\mathrm{off}}$ when the stored energy is almost exhausted. Capturing all wavelets for $t > t_{\mathrm{off}}$ by means of integral (\ref{Basic_A08}) gives the total energy within the capturing surface $S_1$.}
\label{fig:radExtraction_A}
\EF

Note here that the total electromagnetic energy content of the sphere could also be expressed as
\BE
\label{Basic_A09}
\mathcal{W} \left( V_1 , t_\mathrm{off} \right) = \frac{1}{2} \int\limits_{V_1} \left( \mu \left| \boldsymbol{\mathcal{H}} \left( \MAT{r}, t_\mathrm{off} \right) \right|^2 + \epsilon \left| \boldsymbol{\mathcal{E}} \left( \MAT{r}, t_\mathrm{off} \right) \right|^2\right) \D{V}
\EE
which can seem to be simpler than the aforementioned scheme. The simplicity is, however, just formal. The main disadvantage of (\ref{Basic_A09}) is that the integration volume includes also the near-field region, where the fields are rather complex (and commonly singular). Furthermore, contrary to (\ref{Basic_A08}), the radius of the sphere plays an important role in (\ref{Basic_A09}) unlike in (\ref{Basic_A08}), where it appears only via a static time shift $r / c_0$. In fact, it will be shown later on that this dependence can be completely eliminated in the calculation of stored energy.

In order to obtain the stored energy $\WETsto \left( t_\mathrm{off}\right)$ inside $S_1$ we, however, need to know the radiation content of the sphere at $t = t_\mathrm{off}$. A thought experiment aimed at attaining it is presented in figure~\ref{fig:radExtraction}. It exploits the properties of (\ref{Basic_A08}). 
\BF
\includegraphics[width=13.5cm]{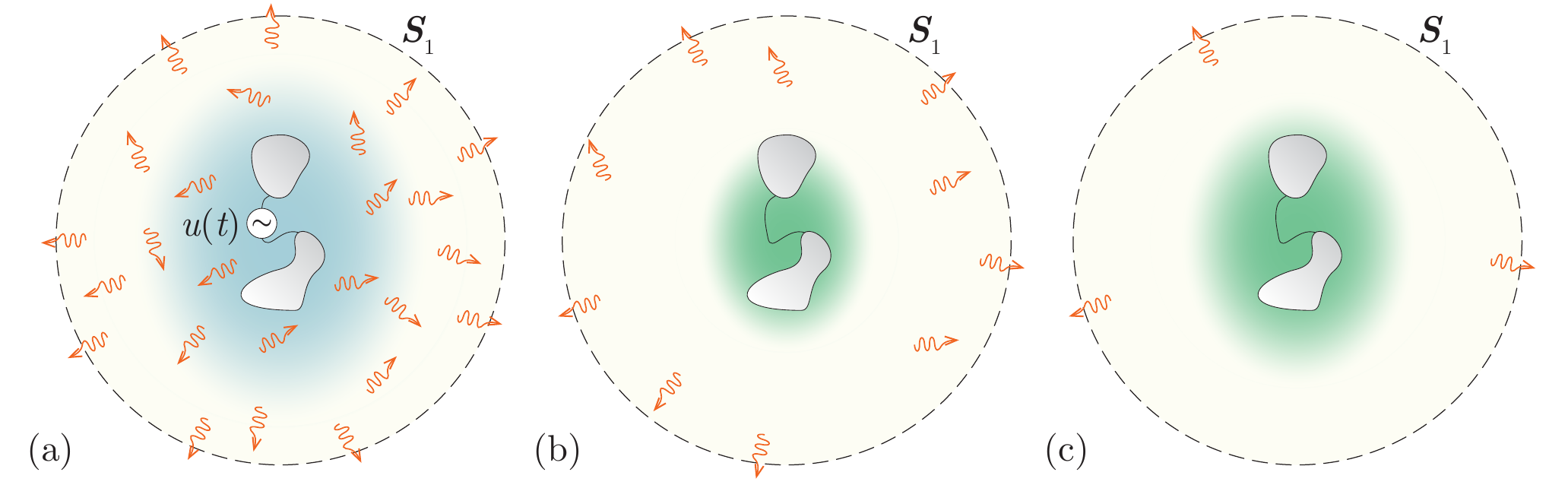}
\caption{Graphical representation of the radiated energy evaluation via (\ref{Basic_A08}) for a loss-less radiator excited by ideal voltage source. Panel (a) shows a steady state just before $t=t_{\mathrm{off}}$, when the steady state radiation (orange wavelets) as well as the steady state stored energy (blue cloud) were maintained by the source. Panel (b) shows that at $t \geq t_{\mathrm{off}}$, the radiating currents are modified so to inhibit any radiation, although they possibly create a new energy storage (green cloud). The radiation emitted before $t=t_{\mathrm{off}}$ (orange wavelets) is unaffected by this modification. Panel (c) depicts the time $t \gg t_{\mathrm{off}}$ when almost all radiation passed $S_1$. The radiation content of the sphere $S_1$ is evaluated via (\ref{Basic_A08}). The green stored energy does not participate as it is not represented by radiation, and is consequently not captured by the integral (\ref{Basic_A08}).}
\label{fig:radExtraction}
\EF
Consulting the figure, let us imagine that during the calculation of $\mathcal{W} \left( V_1, t_\mathrm{off} \right)$ we were capturing the time course of the current $\boldsymbol{\mathcal{J}} \left( \MAT{r} ', t \right)$ at every point. In addition, let us assume that we define an artificial current $\boldsymbol{\mathcal{J}}_{\mathrm{freeze}} \left( \MAT{r} ', t \right)$ as 
\BE
\label{Basic_A0910}
\boldsymbol{\mathcal{J}}_{\mathrm{freeze}} \left( \MAT{r} ', t \right) =  \left\{ \begin{array}{ll}
	\boldsymbol{\mathcal{J}} \left( \MAT{r} ', t \right), & t < t_\mathrm{off}\\
	& \\
	\boldsymbol{\mathcal{J}} \left( \MAT{r} ', t_\mathrm{off} \right), & t \geq t_\mathrm{off}\\
\end{array} \right.
\EE
and use it inside (\ref{Basic_A08}) instead of the true current $\boldsymbol{\mathcal{J}} \left( \MAT{r} ', t \right)$. The expression (\ref{Basic_A08}) then claims that for $t < t_\mathrm{off}$ the artificial current $\boldsymbol{\mathcal{J}}_{\mathrm{freeze}} \left( \MAT{r} ', t \right)$ is radiating in the same way as in the case of  the original problem, but for $t > t_\mathrm{off}$, the radiation is instantly stopped. 
Therefore, if we now evaluate (\ref{Basic_A08}) over the new artificial current, it will give exactly the radiation energy $\mathcal{W}_\mathrm{r} \left( V_1, t_\mathrm{off} \right)$, which has escaped from the sources before $t_\mathrm{off}$. Subtracting it from $\mathcal{W} \left( V_1, t_\mathrm{off} \right)$, we obtain the stored energy $\WETsto \left( t_\mathrm{off}\right)$ and averaging over one period, we obtain the cycle mean stored energy
\BE
\label{Basic_A10}
\WEFsto = \langle\WETsto \left( t_\mathrm{off} \right) \rangle = \frac{1}{T} \int\limits_{\alpha}^{\alpha+T} \WETsto \left(t_\mathrm{off}\right) \D{t_\mathrm{off}}.
\EE
With respect to the freezing of the current, it is important to realize that this could mean an indefinite accumulation of charge at a given point. However, it is necessary to consider this operation as to be performed on the artificial impressed sources, which can be chosen freely.

\BF
\includegraphics[width=12cm]{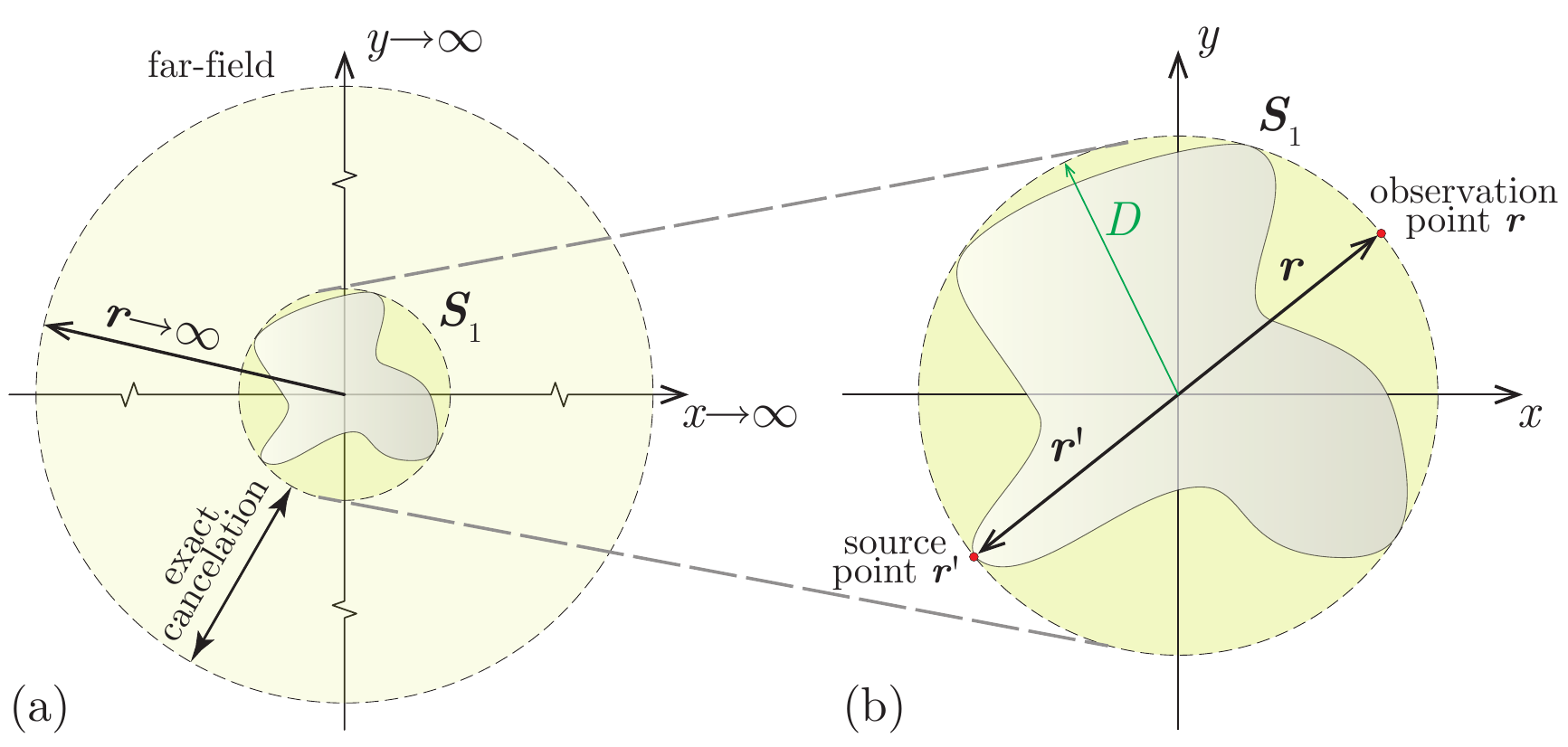}
\caption{Sketch of the far-field cancellation. The circumscribing sphere $\boldsymbol{S}_1$ can be advantageously stretched right around the radiator, since outside this smallest sphere, the first and the second run are identically subtracted.}
\label{fig:farSketch}
\EF

When subtracting the radiated energy from the total energy, it is important to take into account that for $t < t_\mathrm{off}$, the currents were the same in both situations. Thus defining \mbox{$D = \max \{\left| \MAT{r}'\right|\}$}, we can state that for $t < t_\mathrm{off} + \left( r - D \right) / c_0$, the integrals (\ref{Basic_A08}) will exactly cancel during the subtraction, see figure~\ref{fig:farSketch}. The relation (\ref{Basic_A08}) can then be safely evaluated only for $t ' = t - D / c_0 + \MAT{r}_0 \cdot \MAT{r} ' / c_0$ (the worst-case scenario depicted in figure~\ref{fig:farSketch}b), which means that the currents need to be saved only for \mbox{$t > t_\mathrm{off} - 2 D / c_0$}. It is crucial to take into consideration that this is equivalent to say that, after all, the bounding sphere $S_1$ does not need to be situated in the far-field. It is sufficient (and from the computational point of view also advantageous), if $S_1$ is the smallest circumscribing sphere centred in the coordinate system, for the rest of the far-field is cancelled anyhow, see figure~\ref{fig:farSketch}a.

As a final note, we mention that even though the above-described method relies on the integration on a spherical surface, the resulting stored energy properly takes into account the actual geometry of the radiator, representing thus a considerable generalization of the time domain prescription for the stored energy proposed in \cite{Collin_MinimumQofSmallAntennas} which is able to address only the regions outside the smallest circumscribing sphere. Further properties of the method are going to be presented on numerical results in \S\ref{Sec_Antennas} and will be detailed in \S\ref{Sec_Disc}.

\subsection{Stored Energy in Frequency Domain}
\label{FreqDomainEnergyConcept}

This subsection rephrases the stored energy evaluation by Vandenbosch \cite{Vandenbosch_ReactiveEnergiesImpedanceAndQFactorOfRadiatingStructures}, which approaches the issue in the frequency domain, utilizing the complex Poynting's theorem that states \cite{Jackson_ClassicalElectrodynamics} that
\BE
\label{Basics_Eq1}
- \frac{1}{2} \left\langle \MAT{E}, \MAT{J} \right\rangle  = \PMM - \PEE + 2\J\omega\left(\WMM - \WEE \right) = \Pin,
\EE
in which $\Pin$ is the cycle mean complex power, the terms $\PMM$ and $\PEE$ form the cycle mean radiated power $\PMM - \PEE$ and $2\omega\left(\WMM-\WEE\right)$ is the cycle mean reactive net power, and
\BE
\label{Basics_Eq2}
\left\langle \boldsymbol{u}, \boldsymbol{v} \right\rangle = \int\limits_{V} \boldsymbol{u} \left(\MAT{r} \right) \cdot \boldsymbol{v}^\ast \left(\MAT{r} \right) \D{V }
\EE
is the inner product \cite{AkhiezerGlazman_TheoryOfLinearOperatorsInHspace}. In the classical treatment of (\ref{Basics_Eq1}), $\WMM$ and $\WEE$ are commonly taken \cite{Jackson_ClassicalElectrodynamics} as $\mu \left| \MAT{H} \right|^2 / 4$ and $\epsilon \left| \MAT{E} \right|^2 / 4$ that are integrated over the entire space. Both of them are infinite for the radiating system. Nonetheless, when electromagnetic potentials are utilized \cite{Morgenthaler_PowerAndBeautyOfElectromagneticFields}, the complex power balance (\ref{Basics_Eq1}) can be rewritten as
\BE
\label{Basics2_Eq3}
\Pin = \PM-\PE+2\J\omega\left(\WM-\WE\right)  = \frac{\J \omega}{2} \left( \left\langle \MAT{A}, \MAT{J} \right\rangle - \left\langle \varphi, \rho \right\rangle \right),
\EE
where $\MAT{A}$ represents the vector potential, $\varphi$ represents the scalar potential, and $\rho$ stands for the charge density. As an alternative to the classical treatment, it is then possible to write
\BE
\label{Basics_Eq222C_A}
\WM - \J \frac{\PM}{2\omega} = \displaystyle \frac{1}{4} \left\langle \MAT{A} , \MAT{J} \right\rangle
\EE
and
\BE
\label{Basics_Eq222C_B}
\WE - \J \frac{\PE}{2\omega} = \displaystyle \frac{1}{4} \left\langle \varphi, \rho \right\rangle
\EE
without altering (\ref{Basics_Eq1}). However, it is important to stress that in such case, $\WM$ in (\ref{Basics_Eq222C_A}) and $\WE$ in (\ref{Basics_Eq222C_B}) generally represent neither stored nor total magnetic and electric energies \cite{Jackson_ClassicalElectrodynamics}. Some attempts have been undertaken to use (\ref{Basics_Eq222C_A}) and (\ref{Basics_Eq222C_B}) as stored magnetic and electric energies even in non-stationary cases \cite{Carpenter_ElectromagneticEnergyAndPowerInTermsOfChargesAndPotentialsInsteadOfFields}. These attempts were however faced with extensive criticism \cite{Uehara_Allen_Carpenter_ElectromagneticEnergyAndPowerInTermsOfChargesAndPotentialsInsteadOfFields_ANS1}, \cite{Endean_Carpenter_ElectromagneticEnergyAndPowerInTermsOfChargesAndPotentialsInsteadOfFields_ANS1}, mainly due to the variance of separated energies under gauge transformations. 

Regardless of the aforementioned issues, (\ref{Basics_Eq222C_A}) and (\ref{Basics_Eq222C_B}) were modified \cite{Vandenbosch_ReactiveEnergiesImpedanceAndQFactorOfRadiatingStructures} in an attempt to obtain the stored magnetic and electric energies. This modification reads
\begin{subequations}
	\begin{align}
		\label{Basics_Eq3A}
		\WMg &\equiv \WM + \frac{\Wrad}{2}, \\
		\label{Basics_Eq3B}
		\WEg &\equiv \WE + \frac{\Wrad}{2},
	\end{align}
\end{subequations}
where the particular term 
\BE
\label{Basics_Eq4}
\Wrad = \displaystyle\IM\left\{ k \left( k^2 \left\langle {L}_{\mathrm{rad}} \MAT{J}, \MAT{J} \right\rangle - \left\langle {L}_{\mathrm{rad}} \nabla \cdot \MAT{J} , \nabla \cdot \MAT{J} \right\rangle \right) \right\}
\EE
is associated with the radiation field, and the operator
\BE
\label{Basics_Eq5}
{L}_{\mathrm{rad}} \MAT{U} = \frac{1}{16 \pi \epsilon \omega^2} \int\limits_{V '} \MAT{U} \left( \MAT{r} ' \right) \mathrm{e}^{-\J k R} \, \mathrm{d} V '
\EE
is defined using $k = \omega / c_0$ as the wavenumber. The electric currents $\MAT{J}$ are assumed to flow in a vacuum. For computational purposes, it is also beneficial to use the radiation integrals for vector and scalar potentials \cite{Balanis1989}, and rewrite (\ref{Basics_Eq222C_A}), (\ref{Basics_Eq222C_B}) as \cite{CapekJelinekHazdraEichler_MeasurableQ}
\BE
\label{Basics_Eq2C_A}
\WM - \J \frac{\PM}{2\omega} = \displaystyle k^2 \left\langle {L} \MAT{J}, \MAT{J} \right\rangle
\EE
and
\BE
\label{Basics_Eq2C_B}
\WE - \J \frac{\PE}{2\omega} = \displaystyle \left\langle {L}  \nabla \cdot \MAT{J}, \nabla \cdot \MAT{J} \right\rangle,
\EE
with
\BE
\label{Basics_Eq2C_C}
{L} \MAT{U} = \frac{1}{16 \pi \epsilon \omega^2} \int\limits_{V '} \MAT{U} \left( \MAT{r} ' \right) \frac{\mathrm{e}^{-\J k R}}{R} \D{V '}.
\EE
It is suggested in \cite{Vandenbosch_ReactiveEnergiesImpedanceAndQFactorOfRadiatingStructures} that $\WEFstoT = \WMg + \WEg$ is the stored energy $\WEFsto$. Yet this statement cannot be considered absolutely correct, since as it was shown in \cite{GustafssonCismasuJonsson_PhysicalBoundsAndOptimalCurrentsOnAntennas_TAP, JelinekCapekHazdraEichler_LowerBoundOfQz_APS}, $\WEFstoT$ can be negative. Consequently, it is necessary to conclude that $\WEFstoT$, defined by the frequency domain concept \cite{Vandenbosch_ReactiveEnergiesImpedanceAndQFactorOfRadiatingStructures}, can only approximately be equal to the stored energy $\WEFsto$, resulting in
\BE
\label{Basics_Eq6}
\WEFstoT \approx \WEFsto,
\EE
and then by analogy with (\ref{Basic_A01})
\BE
\label{Basics_Eq7}
\widetilde{Q} = 2 \pi \frac{\WEFstoT}{\WETlost} = 2 \pi \frac{\WMg + \WEg}{\WETlost} \approx \QEF
\EE
is defined.

\section{Fractional bandwidth concept of quality factor $Q$}
\label{FBWConcepts}

It is well-known that for \mbox{$Q \gg 1$}, the quality factor $Q$ is approximately inversely proportional to the fractional bandwidth (FBW)
\BE
\label{Basics2_Eq0}
\QFBW \approx \frac{\chi}{\mathrm{FBW}},
\EE
where $\chi$ is a given constant and $\mathrm{FBW} = (\omega^+ - \omega^-) / \omega_0$, \cite{YaghjianBest_ImpedanceBandwidthAndQOfAntennas}. The quality factor $Q$, which is known to fulfil (\ref{Basics2_Eq0}), was found by Yaghjian and Best \cite{YaghjianBest_ImpedanceBandwidthAndQOfAntennas} utilizing an analogy with RLC circuits and using the transition from conductive to voltage standing wave ratio bandwidth. Its explicit definition reads
\BE
\label{Basics2_Eq1}
\QFBW = \frac{\omega }{2\, \RE\{P_{\mathrm{in}}\}} \left|\frac{\partial \Pin}{\partial \omega }\right| = \left| Q_R + \J Q_X \right|,
\EE
where the total input current at the radiator's port is assumed to be normalized to $I_0 = 1$A.

The differentiation of the complex power in the form of (\ref{Basics2_Eq3}) can be used to find the source definition of (\ref{Basics2_Eq1}), and leads to \cite{CapekJelinekHazdraEichler_MeasurableQ}
\begin{subequations}
	\begin{align}
		\label{Basics2_Eq5A}
		Q_R &= \frac{\pi}{\omega} \frac{\PM + \PE + \Prad + \Pom}{ \WETlost}, \\
		\label{Basics2_Eq5B}
		Q_X &= 2\pi \frac{\WEFstoT + \Wom}{\WETlost},
	\end{align}
\end{subequations}
in which
\BE
\label{Basics2_Eq6A}
\frac{\Prad}{2\omega} = \displaystyle \RE\left\{ k \left( k^2 \left\langle {L}_{\mathrm{rad}} \MAT{J} , \MAT{J} \right\rangle - \left\langle {L}_{\mathrm{rad}} \nabla \cdot \MAT{J}, \nabla \cdot \MAT{J} \right\rangle \right) \right\},
\EE
and
\BE
\label{Basics2_Eq6B}
\Wom - \J \frac{\Pom}{2\omega} =  k^2 \left(\left\langle {L} \MAT{J}, D \MAT{J} \right\rangle + \left\langle L \MAT{J}^\ast, D \MAT{J}^\ast \right\rangle \right) 
- \left( \left\langle {L}  \nabla \cdot \MAT{J} , D \nabla \cdot \MAT{J} \right\rangle + \left\langle L \nabla \cdot \MAT{J}^\ast,  D  \nabla \cdot \MAT{J}^\ast  \right\rangle \right).
\EE
The operator $D$ is defined as
\BE
\label{Basics2_Eq7}
D \MAT{U}  = \omega \frac{\partial \MAT{U} }{\partial\omega}.
\EE
As particular cases of (\ref{Basics2_Eq5B}), we obtain the Rhodes' definition \cite{Rhodes_ObservableStoredEnergiesOfElectromagneticSystems} of the quality factor $Q$ as $|Q_X|$ and the definition (\ref{Basics_Eq7}) as $Q_X$, omitting the $\Wom$ term from (\ref{Basics2_Eq5B}).

For the purposes of this paper, we can observe in (\ref{Basic_A01}), (\ref{Basics2_Eq1}), (\ref{Basics2_Eq5A}) and  (\ref{Basics2_Eq5B}) that the stored energy in the case of the FBW concept is equivalent to
\BE
\label{Basics2_Eq10}
\WFBWsto \equiv \frac{1}{2} \left| \frac{\partial \Pin}{\partial\omega} \right| =\left|\WEFstoT + \Wom -\J \frac{\PM + \PE + \Prad + \Pom}{2\omega}\right|,
\EE
but we remark here that (\ref{Basics2_Eq10}) was not intended to be the stored energy \cite{YaghjianBest_ImpedanceBandwidthAndQOfAntennas}.

\section{Non-radiating circuits}
\label{Sec_RLC}

The previous \S\S \ref{EnergyConcepts} and \ref{FBWConcepts} have defined three generally different concepts of stored energy, namely $\WEFsto$, $\WEFstoT$ and $\WFBWsto$. Given that $\WETlost$ is uniquely defined, we can benefit from the  use of the corresponding dimensionless quality factors $\QEF$, $\widetilde{Q}$ and $\QFBW$ for comparing them. This is performed in \S\ref{Sec_RLC} for non-radiating circuits and in \S\ref{Sec_Antennas} for radiating systems. Particularly, in \S\ref{Sec_RLC}, we assume passive lossy but non-dispersive and non-radiating one-ports.

\subsection{Time domain stored energy for lumped elements}
\label{Sec_RLC:TimeDomain}

Following the general procedure indicated in \S\ref{EnergyConcepts}\ref{TimeDomainEnergyConcept}, let us assume a general RLC circuit that was for $t \in \left(-\infty,t_\mathrm{off}\right)$ fed by a time-harmonic source (current or voltage) $s\left( t \right) = \sin \left(\omega_0 t \right)$ which was afterwards switched off for $t \in \left[ t_\mathrm{off},\infty \right)$. Since the circuit is non-radiating, the total energy $\mathcal{W} \left( V_1, t_\mathrm{off}\right)$ is directly equal to $\WETsto \left( t_\mathrm{off} \right)$. Furthermore, a careful selection of the voltage (or current) source for a given circuit helps us to eliminate the internal resistance of the source. So we get
\BE
\label{RLC_time4}
\WEFsto = \sum\limits_k \frac{R_k}{T} \int\limits_{\alpha}^{\alpha+T} \int\limits_{t_\mathrm{off}}^{\infty} i_{\mathrm{R},k}^2 \left( t \right) \D{t} \D{t_\mathrm{off}},
\EE
where $i_{\mathrm{R},k} \left( t \right)$ is the transient current in the $k$-th resistor.

The currents $i_{\mathrm{R},k}$ are advantageously evaluated in the frequency domain. The Fourier transform of the source reads \cite{Rothwell_Cloud_Electromagnetics}
\BE
\label{RLC_time_1}
S\left(\omega\right) = \frac{\J \pi}{2} \left( \delta\left(\omega+\omega_0\right) - \delta\left(\omega-\omega_0\right) \right) + \frac{\mathrm{e}^{-\J \omega t_\mathrm{off}}}{2} \left( \frac{\mathrm{e}^{\J \omega_0 t_\mathrm{off}}}{\omega - \omega_0} - \frac{\mathrm{e}^{-\J \omega_0 t_\mathrm{off}}}{\omega + \omega_ 0}\right).
\EE
We can then write \mbox{$I_{\mathrm{R},k} \left(\omega\right) = T_{\mathrm{R}_k}\left(\omega\right) S\left(\omega\right)$}, where $T_{\mathrm{R}_k} \left(\omega\right)$ represents the transfer function. Consequently
\BE
\begin{split}
	\label{RLC_time_2}
	i_{\mathrm{R},k}\left( t \right) &= \frac{1}{2 \pi} \int\limits_{-\infty}^{\infty} T_{\mathrm{R}_k}\left(\omega\right) S\left(\omega\right) \mathrm{e}^{\J \omega t} \D{\omega}  \\
	& =  \frac{1}{2} \IM\left\{T_{\mathrm{R}_k}\left(\omega_0\right) \mathrm{e}^{\J \omega_0 t} \right\} + \frac{\omega}{4 \pi} \int\limits_{-\infty}^{\infty} T_{\mathrm{R}_k}\left(\omega\right) \left( \frac{\mathrm{e}^{\J \omega_0 t_\mathrm{off}}}{\omega - \omega_0} - \frac{\mathrm{e}^{-\J \omega_0 t_\mathrm{off}}}{\omega + \omega_0} \right)  \mathrm{e}^{\J \omega \left(t- t_\mathrm{off}\right)} \D{\omega}.
\end{split}
\EE
As the studied circuit is lossy, $T_{\mathrm{R}_k} \left(\omega\right)$ has no poles on the real $\omega$-axis and the second integral can be evaluated by the standard contour integration in the complex plane of $\omega$ along the semi-circular contour in the upper $\omega$ half-plane, while omitting the points $\omega = \pm \omega_0$. The result of the contour integration for $t > t_\mathrm{off}$ can be written as
\BE
\label{RLC_time3}
i_{\mathrm{R},k}\left( t \right) = \frac{\J}{2} \sum\limits_m  \mathop{\mathrm{res}}\limits_{\omega \rightarrow \omega_{m,k}} \Bigg\{ T_{\mathrm{R}_k} \left(\omega\right) \Bigg(\frac{\mathrm{e}^{\J \omega_0 t_\mathrm{off}}}{\omega - \omega_0} - \frac{\mathrm{e}^{- \J \omega_0 t_\mathrm{off}}}{\omega + \omega_0} \Bigg) \mathrm{e}^{\J\omega \left(t- t_\mathrm{off}\right)} \Bigg\},
\EE
where $\omega_{m,k}$ are the poles of $T_{\mathrm{R}_k}\left(\omega\right)$ with $\IM\left\{\omega_{m,k}\right\} > 0$. The substitution of (\ref{RLC_time3}) into (\ref{RLC_time4}) gives the mean stored energy. It is also important to realize that in this case, it is easy to analytically carry out both integrations involved in (\ref{RLC_time4}). The result is obviously identical to the cycle mean of the classical definition of stored energy.
\BE
\label{RLC_time5}
\WETsto \left( t_\mathrm{off} \right) = \frac{1}{2} \left(\sum\limits_m L_m i_{\mathrm{L},m}^2 \left(t_\mathrm{off} \right) + \sum\limits_n C_n u_{\mathrm{C},n}^2 \left( t_\mathrm{off} \right) \right),
\EE
which is the lumped circuit form of (\ref{Basic_A09}), with $i_{\mathrm{L},m} \left( t \right)$ being the current in the $m$-th inductor $L_m$ and $u_{\mathrm{C},n} \left( t \right)$ being the voltage on the $n$-th capacitor $C_n$.

\subsection{Frequency domain stored energy for lumped elements}
\label{Sec_RLC:FreqDomain}

Without the radiation (\mbox{$\Prad = 0, \omega\Wrad = 0$}), the cycle mean of (\ref{Basic_A09}), which is also equal to the cycle mean (\ref{RLC_time5}), is identical to the frequency domain expression
\BE
\label{RLC_G1}
\WEFstoT = \WM + \WE = \frac{1}{4} \left(\sum\limits_m L_m |I_{\mathrm{L},m}|^2 + \sum\limits_n C_n |U_{\mathrm{C},n}|^2 \right) = \frac{1}{4} \int\limits_{V} \left( \mu \left|\MAT{H}\right|^2 + \epsilon \left|\MAT{E} \right|^2\right)\D{V},
\EE
where $\WM$ and $\WE$ are defined by (\ref{Basics_Eq2C_A}) and (\ref{Basics_Eq2C_B}) respectively.
We thus conclude that $\WEFsto = \WEFstoT$ and $\QEF = \widetilde{Q}$ for non-radiating circuits.

\subsection{Frequency domain stored energy for lumped elements derived from FBW concept}
\label{Sec_RLC:FreqDomainFBW}

In order to evaluate (\ref{Basics2_Eq1}), the same procedure as in the derivation of Foster's reactance theorem \cite{Harrington_TimeHarmonicElmagField} can be employed (keeping in mind the unitary input current, no radiation and assuming non-zero conductivity). It results in
\BE
\begin{split}
	\label{RLC_2}
	W_\mathrm{sto}^\mathrm{FBW} &= \left| \frac{1}{4} \int\limits_{V} \left( \mu \left|\MAT{H}\right|^2 + \epsilon \left|\MAT{E} \right|^2\right)\D{V} - \frac{\J \sigma}{2}\int\limits_{V} \MAT{E}^\ast\cdot\frac{\partial\MAT{E}}{\partial\omega}\D{V} \right| \\
	&= \Bigg| \frac{1}{4} \left( \sum\limits_m L_m |I_{\mathrm{L},m}|^2 + \sum\limits_n C_n |U_{\mathrm{C},n}|^2 \right) - \frac{\J}{2} \sum\limits_k R_k I_{\mathrm{R},k}^\ast \frac{\partial I_{\mathrm{R},k}}{\partial\omega}\Bigg|,
\end{split}
\EE
where $I_{\mathrm{R},k}$ is the amplitude of the current through the $k$-th resistor. The formula indicated above clearly reveals the fundamental difference between $\WEFsto$ and $\WFBWsto$, which consists in the last term of RHS in (\ref{RLC_2}). It means that, in general, $\WFBWsto$ does not represent the time-averaged stored energy.

\subsection{Results}
\label{Sec_RLC:Results}

In the previous \S\S\ref{Sec_RLC}\ref{Sec_RLC:TimeDomain}--\ref{Sec_RLC}\ref{Sec_RLC:FreqDomainFBW} we have shown that for non-radiating circuits there is no difference between the quality factor defined in the time domain ($\QEF$) and the one defined in the frequency domain ($\widetilde{Q}$). Nevertheless, there is a substantial difference between $\QEF$ and $\QFBW$, which is going to be presented in \S\ref{Sec_RLC}\ref{Sec_RLC:Results} using two representative examples depicted in figure~\ref{fig:RLC}.
\BF
\includegraphics[width=12cm]{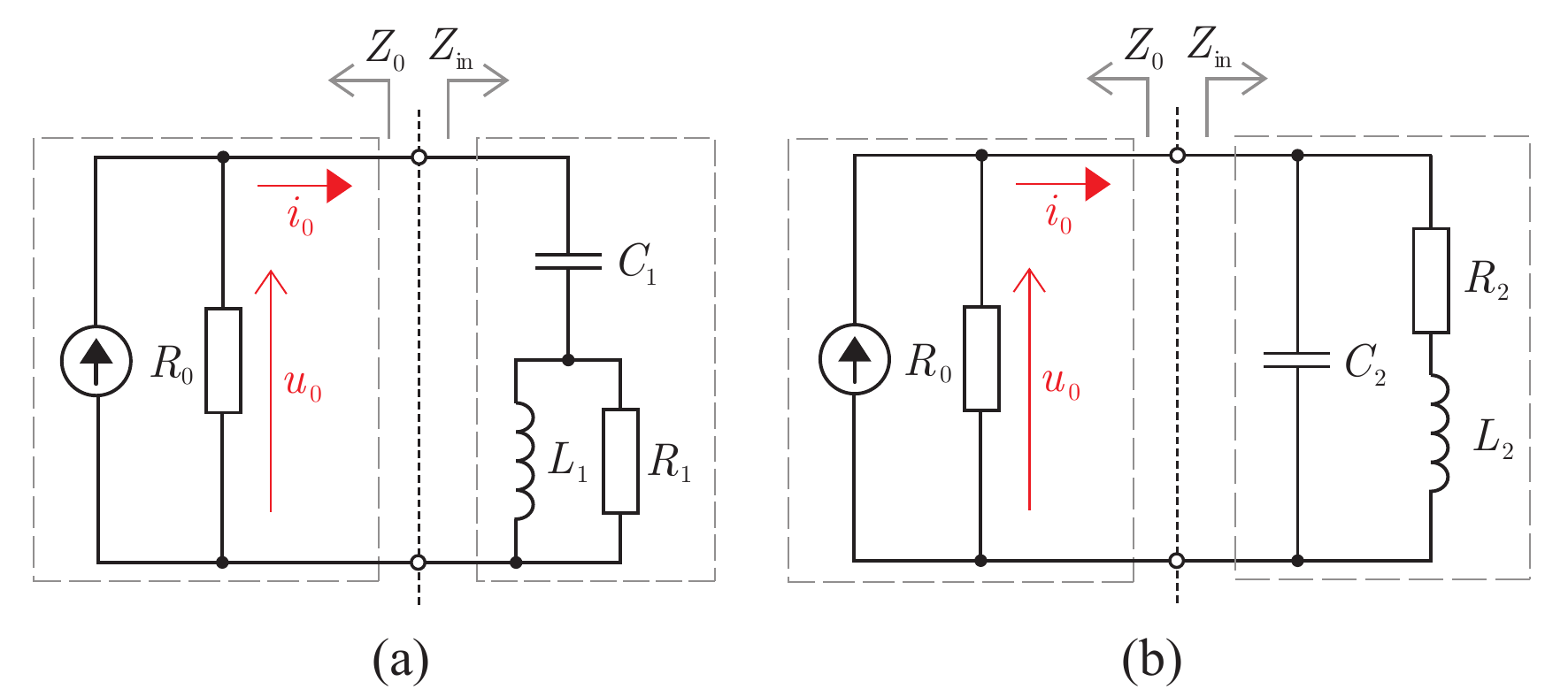}
\caption{Studied RLC circuits: (a) $C_1$ in series with parallel $L_1$ and $R_1$, and (b) $C_2$ in parallel with serial $R_2$ and $L_2$}. 
\label{fig:RLC}
\EF
We do not explicitly consider simple series and parallel RLC circuits in this paper, since the three definitions of the stored energy and quality factor $Q$ deliver exactly the same results, i.e. \mbox{$Q = \omega_0 L / R = \omega_0 R C$}. This is attributable to the frequency independence of the current flowing through the resistor (the series resonance circuit), or of the voltage on the resistor (the parallel resonance circuit). In those cases, the last term of (\ref{RLC_2}) vanishes identically. This fact is the very reason why the FBW approach works perfectly for radiators that can be approximated around resonance by a parallel or series RLC circuit. However, it also means that for radiators that need to be approximated by other circuits, the approach may not deliver the correct energy. This is probably the reason why this method seems to fail in the case of wideband radiators and radiators with slightly separated resonances.

In the case of circuits depicted in figure~\ref{fig:RLC}, the input impedances are
\BE
\label{RLC_ex1}
Z_\mathrm{in}^{(a)} = \displaystyle\frac{1}{\J\omega C_1} + \frac{1}{\displaystyle\frac{1}{R_1} + \frac{1}{\J\omega L_1}}, \quad Z_\mathrm{in}^{(b)} = \displaystyle\frac{1}{\displaystyle\frac{1}{R_2 + \J\omega L_2} + \J\omega C_2},
\EE
and the corresponding resonance frequencies read
\BE
\label{RLC_ex2}
\omega_0^{(a)} = \frac{R_1}{L_1}\frac{1}{\displaystyle\sqrt{\frac{C_1 R_1^2}{L_1}-1}},\quad\omega_0^{(b)} = \frac{R_2}{L_2}\displaystyle\sqrt{\frac{L_2}{C_2 R_2^2}-1},
\EE
respectively. Utilizing the method from \S\ref{EnergyConcepts}\ref{Sec_RLC:TimeDomain}, it can be demonstrated that the energy quality factors are
\BE
\label{RLC_ex4}
\QEF^{(a)} = \widetilde{Q}^{(a)} = \frac{R_1}{\omega_0^{(a)} L_1}, \quad \QEF^{(b)} = \widetilde{Q}^{(b)} = \frac{\omega_0^{(b)} L_2}{R_2},
\EE
while the FBW quality factors equal
\BE
\label{RLC_ex5}
\QFBW^{(a)} = \kappa^{(a)} \QEF^{(a)}, \quad \QFBW^{(b)} = \kappa^{(b)} \QEF^{(b)},
\EE
where
\BE
\label{RLC_ex51}
\kappa^{(a)} = \frac{1}{\omega_0^{(a)} \sqrt{L_1 C_1}}, \quad \kappa^{(b)} = \omega_0^{(b)} \sqrt{L_2 C_2}.
\EE
For the sake of completeness, it is useful to indicate that the quality factors proposed by Rhodes \cite{Rhodes_ObservableStoredEnergiesOfElectromagneticSystems} are found to be
\BE
\label{RLC_ex6}
\left|Q_X^{(a)} \right| = \left(\kappa^{(a)} \right)^2 \QET^{(a)}, \quad \left| Q_X^{(b)} \right| = \left(\kappa^{(b)} \right)^2 \QET^{(b)}.
\EE

The comparison of the above-mentioned quality factors is depicted in figure~\ref{fig:RLC2} using the parametrization by $R_i / L_i$ and $R_i C_i$, where $i \in \left\{1, 2\right\}$. The circuit (a) in figure~\ref{fig:RLC} is resonant for \mbox{$R_1 C_1 > L_1 / R_1$}, whilst the circuit (b) in the same figure is resonant for \mbox{$R_1 C_1 < L_1 / R_1$}. It can be observed that the difference between the depicted quality factors decreases as the quality factor rises and finally vanishes for $Q \rightarrow \infty$. On the other hand, there are significant differences for $Q < 2$.
\BF
\includegraphics[width=\figwidth cm]{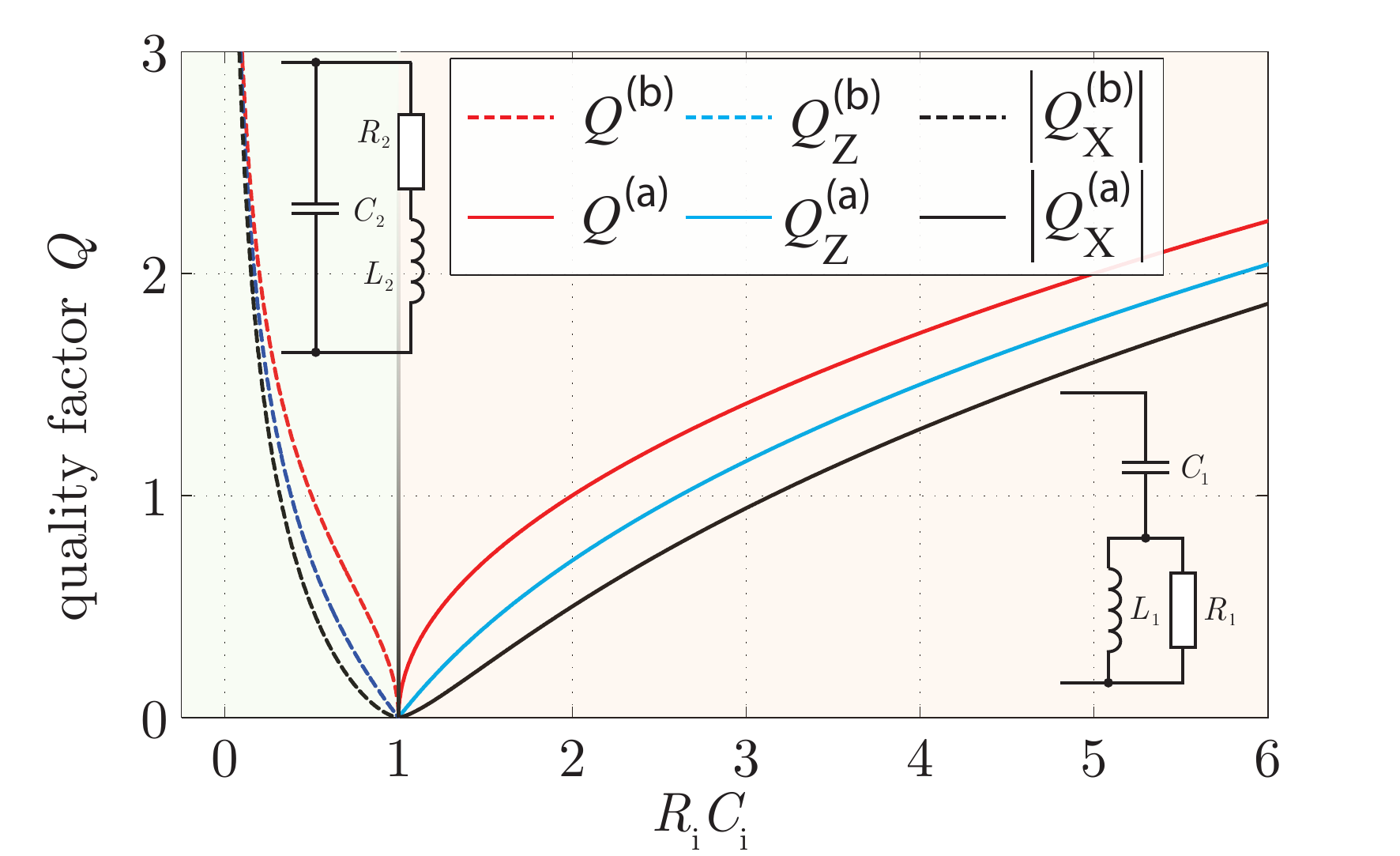}
\caption{The quality factors for the two particular lumped RLC circuits of figure~\ref{fig:RLC}. The curves correspond to \mbox{$R_i / L_i = 1 \,\mathrm{s}^{-1}$}, with $i \in \{1,2\}$.}
\label{fig:RLC2}
\EF

Therefore, we can conclude that for general RLC circuits made of lumped (non-radiating) elements
\BE
\label{RLC_ex7}
\WEFsto \equiv \WEFstoT \neq \WFBWsto \Longrightarrow \QEF \equiv \widetilde{Q} \neq \QFBW.
\EE

\section{Radiating structures}
\label{Sec_Antennas}

The evaluation of the quality factor $Q$ for radiating structures is far more involved than for non-radiating circuits. This is due to the fact that the radiating energy should be subtracted correctly. Hence, the method proposed in \S\ref{EnergyConcepts}\ref{TimeDomainEnergyConcept} was implemented according to the flowchart depicted in figure~\ref{fig:radStructures}.

The evaluation is done in Matlab \cite{matlab}. The current density $\boldsymbol{\mathcal{J}} \left(\boldsymbol{r} ' ,t\right)$ and the current $i_{\mathrm{R}_0} \left(t\right)$ flowing through the internal resistance of the source are the only input quantities used, see figure~\ref{fig:radStructures}. In particular cases treated in this section, we utilize the ideal voltage source that invokes $i_{\mathrm{R}_0} \left(t\right) \equiv 0$, and thus the first integral in RHS of (\ref{Basic_A05}) vanishes.
\BF
\includegraphics[width=13cm]{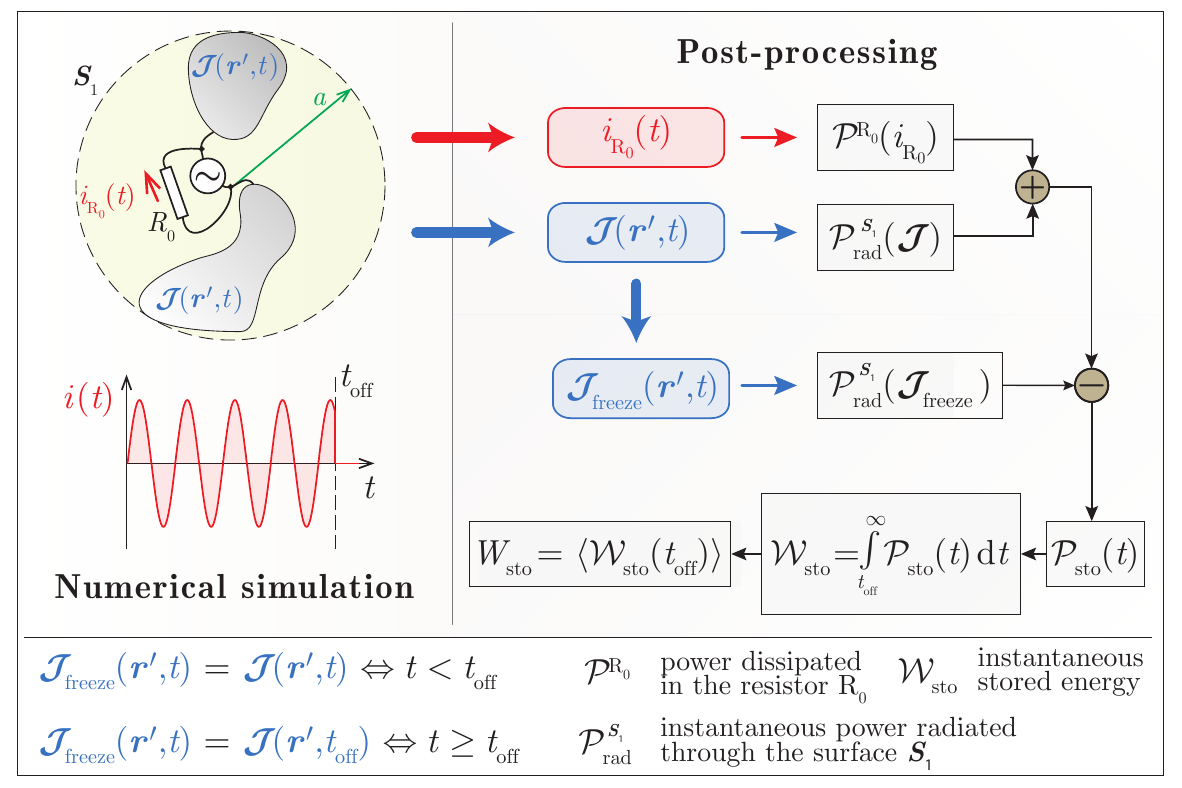}
\caption{Flowchart of the method proposed in \S\ref{EnergyConcepts}\ref{TimeDomainEnergyConcept}. The time domain currents are processed according to the right-hand side of the flowchart.}
\label{fig:radStructures}
\EF

In order to verify the proposed approach, three types of radiators are going to be calculated, namely the centre-fed dipole, off-centre-fed dipole and Yagi-Uda antenna. All these radiators are made of an infinitesimally thin-strip perfect electric conductor and operate in vacuum background. Consequently, the second integral in RHS of (\ref{Basic_A05}) also vanishes. The quality factor $Q$ calculated with the help of the novel method is going to be compared with the results of two remaining classical approaches detailed in \S\ref{EnergyConcepts}\ref{FreqDomainEnergyConcept} and \S\ref{FBWConcepts}, which produced the quality factors $\widetilde{Q}$ (\ref{Basics_Eq7}) and $\QFBW$ (\ref{Basics2_Eq1}) respectively. 

All essential steps of the method are going to be explained using the example of a centre-fed dipole in \S\ref{Sec_Antennas}\ref{Sec_Antennas:Dipole_Ex1}. Subsequently, in \S\ref{Sec_Antennas}\ref{Sec_Antennas:offCentreDipole_Ex2} and \S\ref{Sec_Antennas}\ref{Sec_Antennas:YagiUda_Ex3}, the method is going to be directly applied to more complicated radiators. The most important properties of the novel method are going to be examined in the subsequent discussion \S\ref{Sec_Disc}.

\subsection{Centre-fed thin-strip dipole}
\label{Sec_Antennas:Dipole_Ex1}

The first structure to be calculated is a canonical radiator: a dipole of the length $L$ and width \mbox{$w = L/200$}. The dipole is fed by a voltage source \cite{Balanis1989} located in its centre.

The calculation starts in FEKO commercial software \cite{feko} in which the dipole is simulated. The dipole is fed by a unitary voltage and the currents \mbox{$\boldsymbol{J} \left(\boldsymbol{r} ', \omega\right)$} are evaluated within the frequency span from $ka = 0$ to $ka \approx 325$ for 8192~samples. The resulting currents are imported into Matlab. We define the normalized time \mbox{$t_\mathrm{n} = t \omega_0 / \left( 2 \pi \right)$} (see x-axis of figures~\ref{fig:DipoleFig1} and \ref{fig:DipoleFig2}), where $\omega_0$ is the angular frequency that the quality factor $Q$ is going to be calculated at. Then iFFT over $S \left(\omega\right) \boldsymbol{J} \left(\boldsymbol{r} ', \omega\right) $, see (\ref{RLC_time_1}), is applied, and the time domain currents \mbox{$\boldsymbol{\mathcal{J}} \left(\boldsymbol{r}',t\right)$} with \mbox{$\Delta t_n = 0.02$} for $t_n \in \left(0 , 163\right) $ are obtained. The implementation details of iFFT, which must also contain singularity extraction of the source spectrum $S \left(\omega\right)$, are not discussed here, as they are not of importance to the method of quality factor calculation itself. The next step consists in the evaluation of (\ref{Basic_A08}) for both, the original currents \mbox{$\boldsymbol{J} \left(\boldsymbol{r} ', t\right)$} and frozen currents \mbox{$\boldsymbol{J}_\mathrm{freeze} \left(\boldsymbol{r} ', t\right)$}, see figure~\ref{fig:radStructures}.
\BF
\includegraphics[width=\figwidth cm]{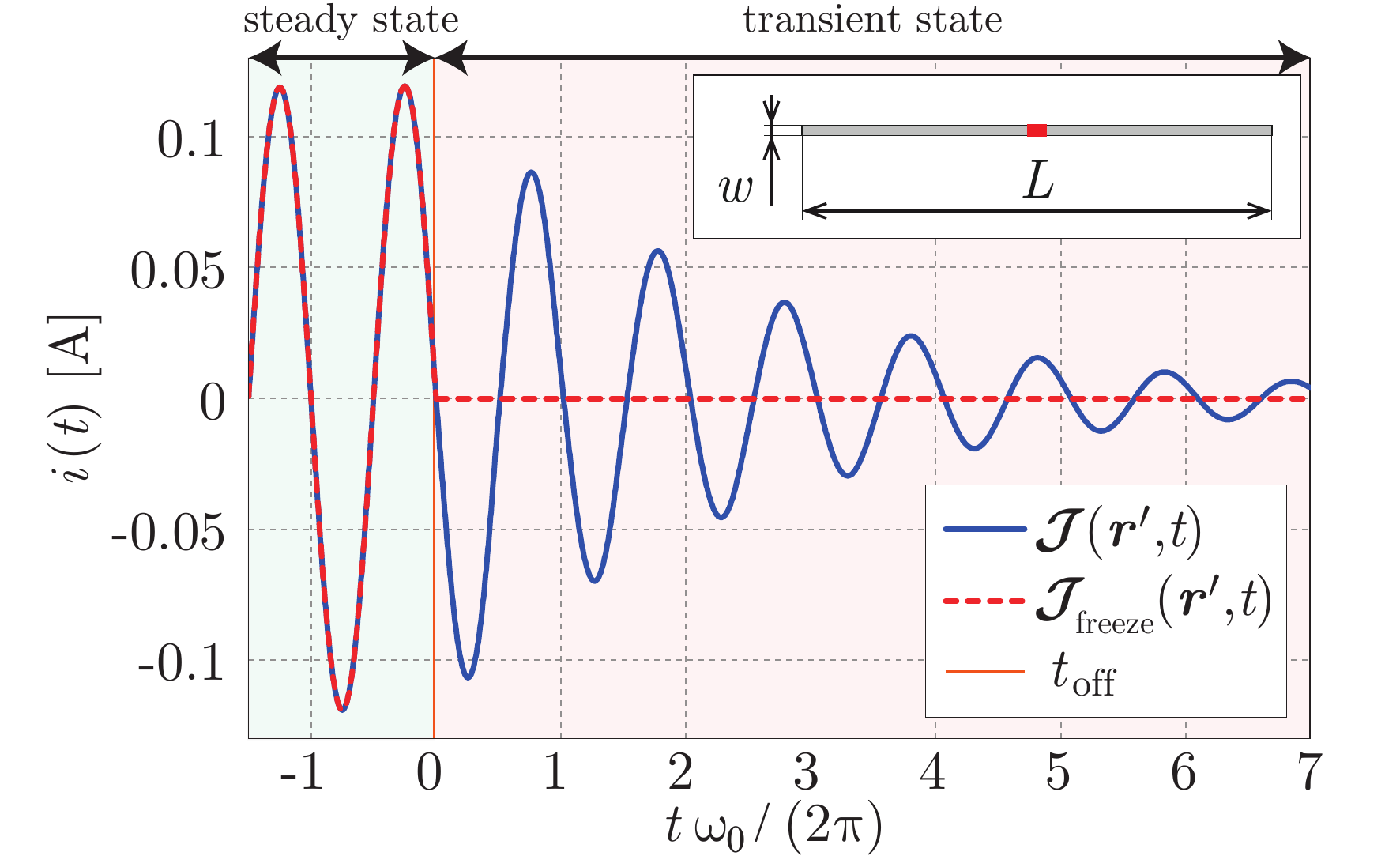}
\caption{Current flowing through the voltage gap of dipole (exact proportions of the antenna are indicated in the inset). The blue line shows the steady state and the transient state of the original currents $\boldsymbol{\mathcal{J}} \left(\boldsymbol{r}',t \right)$, whilst the red line corresponds to the modified currents \mbox{$\boldsymbol{\mathcal{J}}_\mathrm{freeze}\left( \MAT{r}',t\right)$}. The depicted curves correspond to the source with the voltage \mbox{$u\left( t \right) = U_0 \sin\left(\omega_0 t \right) H\left(t_\mathrm{off} - t\right)$}, where the $U_0$ was chosen so that the mean radiated power equals $0.5$\,W.}
\label{fig:DipoleFig1}
\EF
\BF
\includegraphics[width=\figwidth cm]{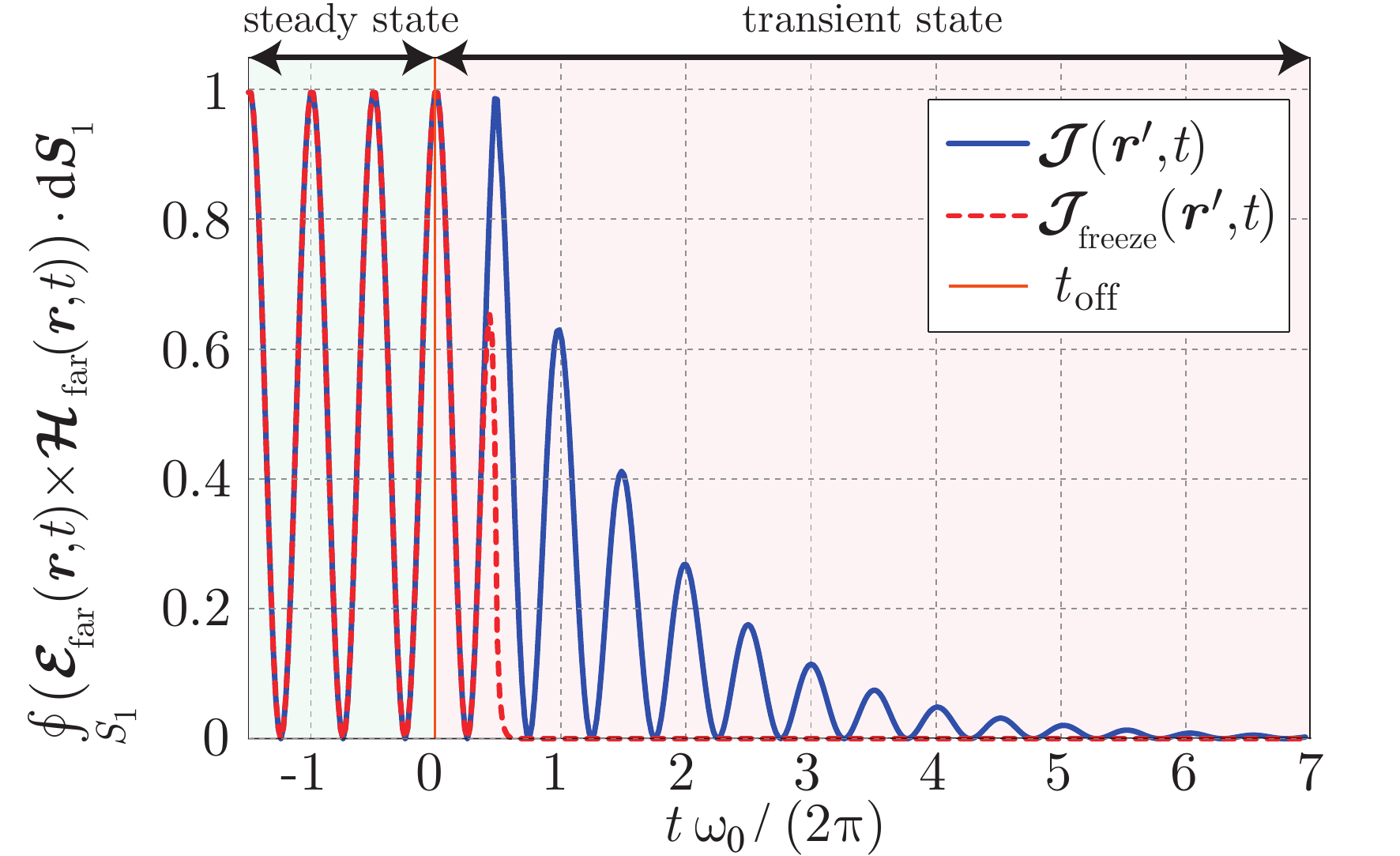}
\caption{Radiated power passing through the surface $\boldsymbol{S}_1$ for a centre-fed dipole. The meaning of the blue and red lines as well as the normalization of input voltage is the same as in figure~\ref{fig:DipoleFig1}.}
\label{fig:DipoleFig2}
\EF

At this point, it is highly instructive to explicitly show the time course of the current at the centre of the dipole (see figure~\ref{fig:DipoleFig1}), as well as the time course  of the power passing through the surface $S_1$ in both aforementioned scenarios (original and frozen currents), see figure~\ref{fig:DipoleFig2}. The source was switched off at \mbox{$t_\mathrm{n} = 0$}. During the following transient (blue lines in figure~\ref{fig:DipoleFig1} and figure~\ref{fig:DipoleFig2}), all energy content of the sphere is lost by the radiation. Within the second scenario, with all currents constant for $t \geq t_\mathrm{off}$, the radiation of the dipole is instantaneously stopped at $t_\mathrm{off} = 0$. The power radiated for \mbox{$t_\mathrm{off}>0$} (red line in figure~\ref{fig:DipoleFig2}) then represents the radiation that existed at $t = t_\mathrm{off}$ within the sphere, but needed some time to leave the volume. Subtracting the blue and red curves in figure~\ref{fig:DipoleFig2} and integrating in time for \mbox{$t \geq t_\mathrm{off}$} then gives the stored energy at \mbox{$t = t_\mathrm{off}$}. In order to construct the course of $\QET \left(t_\mathrm{off}\right)$, the stored energy is evaluated for six different switch-off times $t_\mathrm{off}$. The resulting $\QET\left(t_\mathrm{off}\right)$ is then fitted by 
\BE
\label{Dip_3}
\QET\left(t_\mathrm{off}, \omega_0\right) = A + B \sin\left(2\omega_0 t_\mathrm{off}+\beta\right).
\EE
The fitting was exact (within the used precision) in all fitted points, which allowed us to consider (\ref{Dip_3}) as an exact expression for all $t_\mathrm{off}$. The constant $A$ then equals $Q \left(\omega_0\right)$. 

We are typically interested in the course of $Q$ with respect to the frequency. Repeating the above-explained procedure for varying $\omega_0$, we obtain the red curve in figure~\ref{fig:DipoleFig3}. In the same figure, the comparison with $\widetilde{Q}$ from \S\ref{EnergyConcepts}\ref{FreqDomainEnergyConcept} (blue curve) and $\QFBW$ from \S\ref{FBWConcepts} (green curve) is depicted. To calculate $\widetilde{Q}$ by means of (\ref{Basics_Eq2C_A}), (\ref{Basics_Eq2C_B}), (\ref{Basics_Eq4}), (\ref{Basics_Eq6}) in the frequency domain, we used the currents \mbox{$\boldsymbol{J} \left(\boldsymbol{r} ', \omega\right)$} from FEKO and renormalized them with respect to the input current $I_0 = 1$\,A. Similarly, the calculation of the FBW quality factor $\QFBW$ (\ref{Basics2_Eq1}) is performed for the same source currents with identical normalization, and is based on expression (\ref{Basics2_Eq10}) and all subsequent relations integrated in Matlab.
\BF
\includegraphics[width=\figwidth cm]{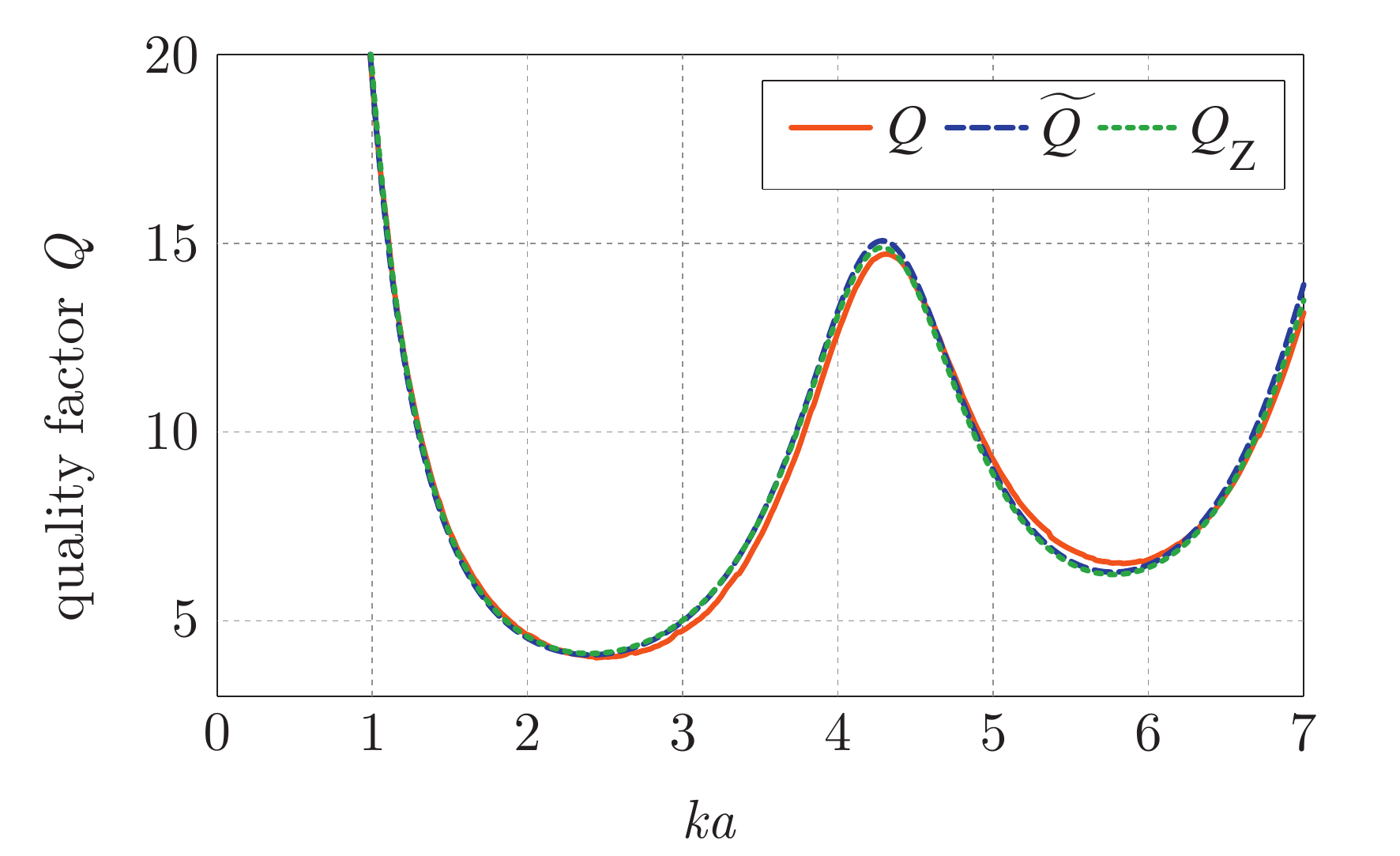}
\caption{Frequency dependence of the quality factors for a centre-fed dipole.}
\label{fig:DipoleFig3}
\EF

\subsection{Off-centre-fed thin-strip dipole}
\label{Sec_Antennas:offCentreDipole_Ex2}

The second example is represented by an off-centre-fed dipole, which is known to exhibit the zero value of $\QFBW$ \cite{GustafssonJonsson_AntennaQandStoredEnergiesFieldsCurrentsInputImpedance}, for \mbox{$ka \approx 6.2$} provided that the delta gap is placed at $0.23 L$ from the bottom of the dipole. The dipole has the same parameters as in the previous example, except the position of feeding (see the inset in figure~\ref{fig:OffCentreDipoleFig1}).
\BF
\includegraphics[width=\figwidth cm]{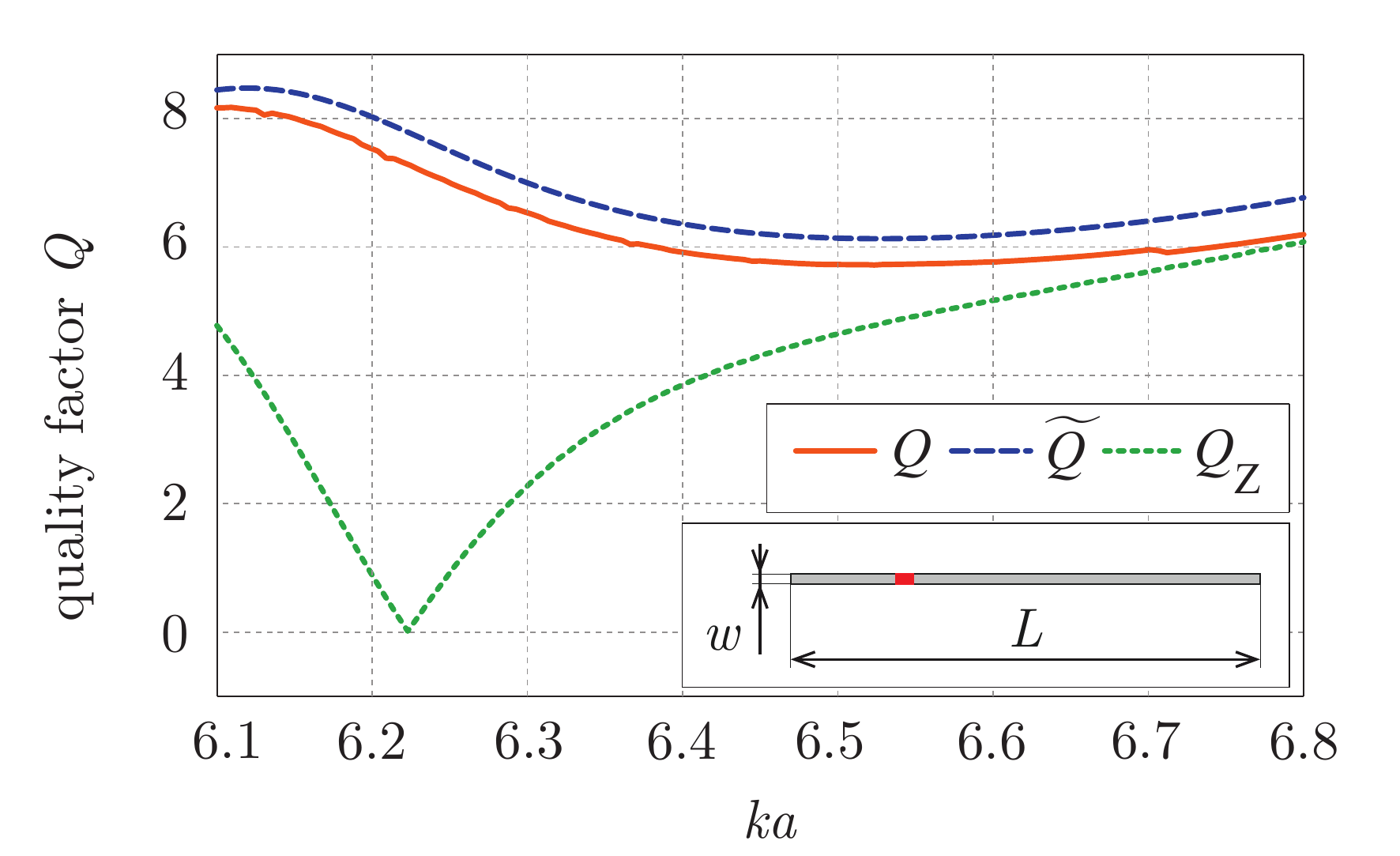}
\caption{Comparison of all quality factors in the case of off-centre-fed dipole (see the inset for exact proportions of the antenna).}
\label{fig:OffCentreDipoleFig1}
\EF

It is apparent from figure~\ref{fig:OffCentreDipoleFig1} that the quality factor $Q$ based on the new stored energy evaluation does not suffer from drop-off around $ka \approx 6.2$, and in fact yields similar values as $\widetilde{Q}$, including the same trend.

\subsection{Yagi-Uda antenna}
\label{Sec_Antennas:YagiUda_Ex3}

Yagi-Uda antenna was selected as a representative of quite complex structure that the method can ultimately be tested on. The antenna has the same dimensions as in \cite{YaghjianBest_ImpedanceBandwidthAndQOfAntennas} and is depicted in the inset in figure~\ref{fig:YagiFig2}. Since this antenna has non-unique phase centre, it can serve as an ideal candidate for verification of the coordinate independence of the novel method.
\BF
\includegraphics[width=\figwidth cm]{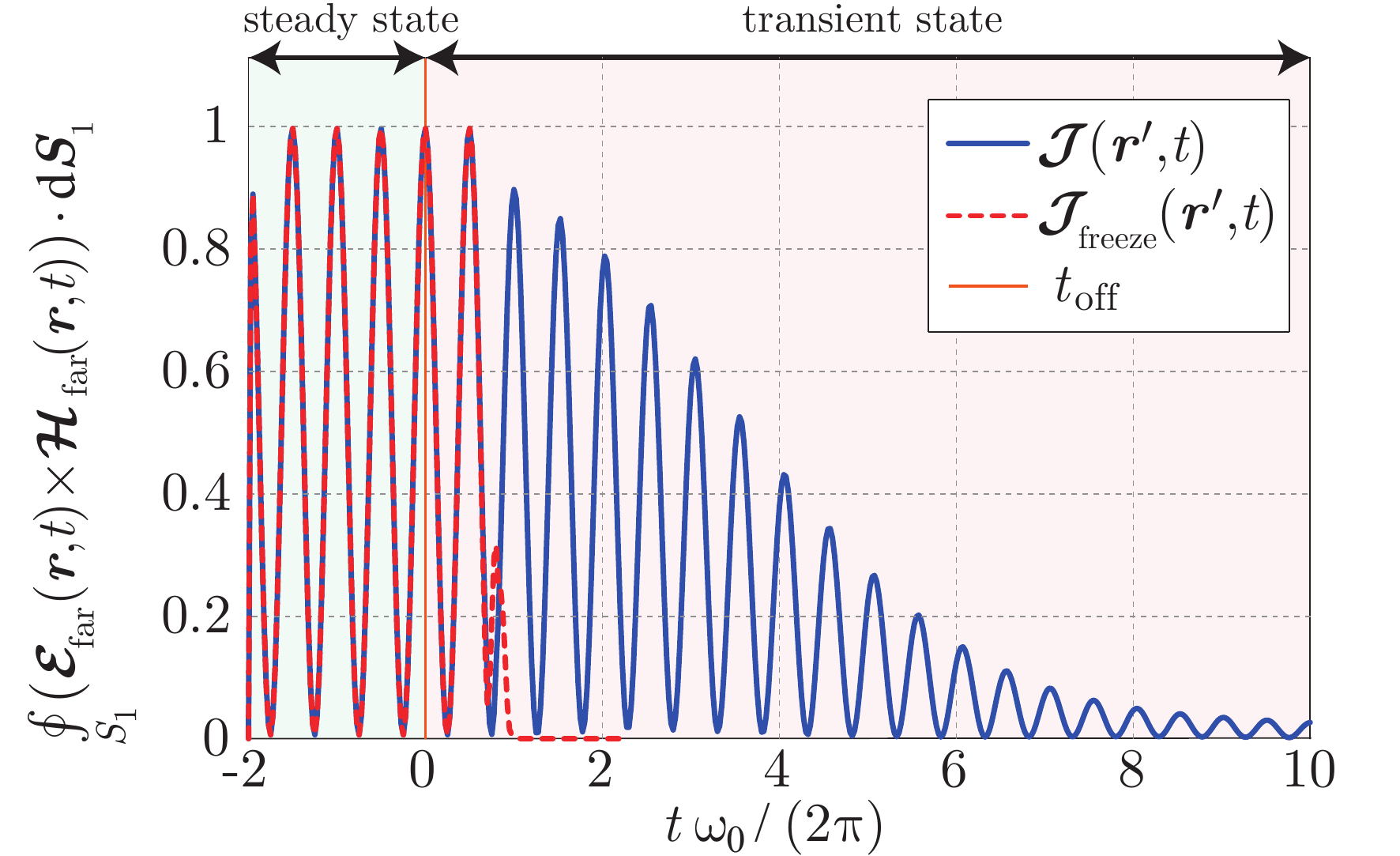}
\caption{Radiated power passing through the surface $\boldsymbol{S}_1$ in the case of Yagi-Uda antenna. The meaning of the blue and red lines as well as the normalization of input voltage is the same as in figure~\ref{fig:DipoleFig1}. The antenna proportions are depicted in the inset of figure~\ref{fig:YagiFig2}.}
\label{fig:YagiFig1}
\EF
The results were calculated in the same way as in the previous examples, and are indicated in figures~\ref{fig:YagiFig1} and \ref{fig:YagiFig2}. The comparison between the results in figure~\ref{fig:YagiFig1} and those related to the dipole in figure~\ref{fig:DipoleFig2} clearly reveals that the transient state is remarkably longer in the case of Yagi-Uda antenna, which means that the longer integration time is required. Furthermore, it can be seen (red curve for $t > t_\mathrm{off}$) that the bounding sphere contains a considerable amount of radiation that should be subtracted. The accuracy of this subtraction is embodied in figure~\ref{fig:YagiFig2}, which shows the quality factors $Q$, $\widetilde{Q}$, and $\QFBW$. Notice the similarity between $Q$ and $\widetilde{Q}$.
\BF
\includegraphics[width=\figwidth cm]{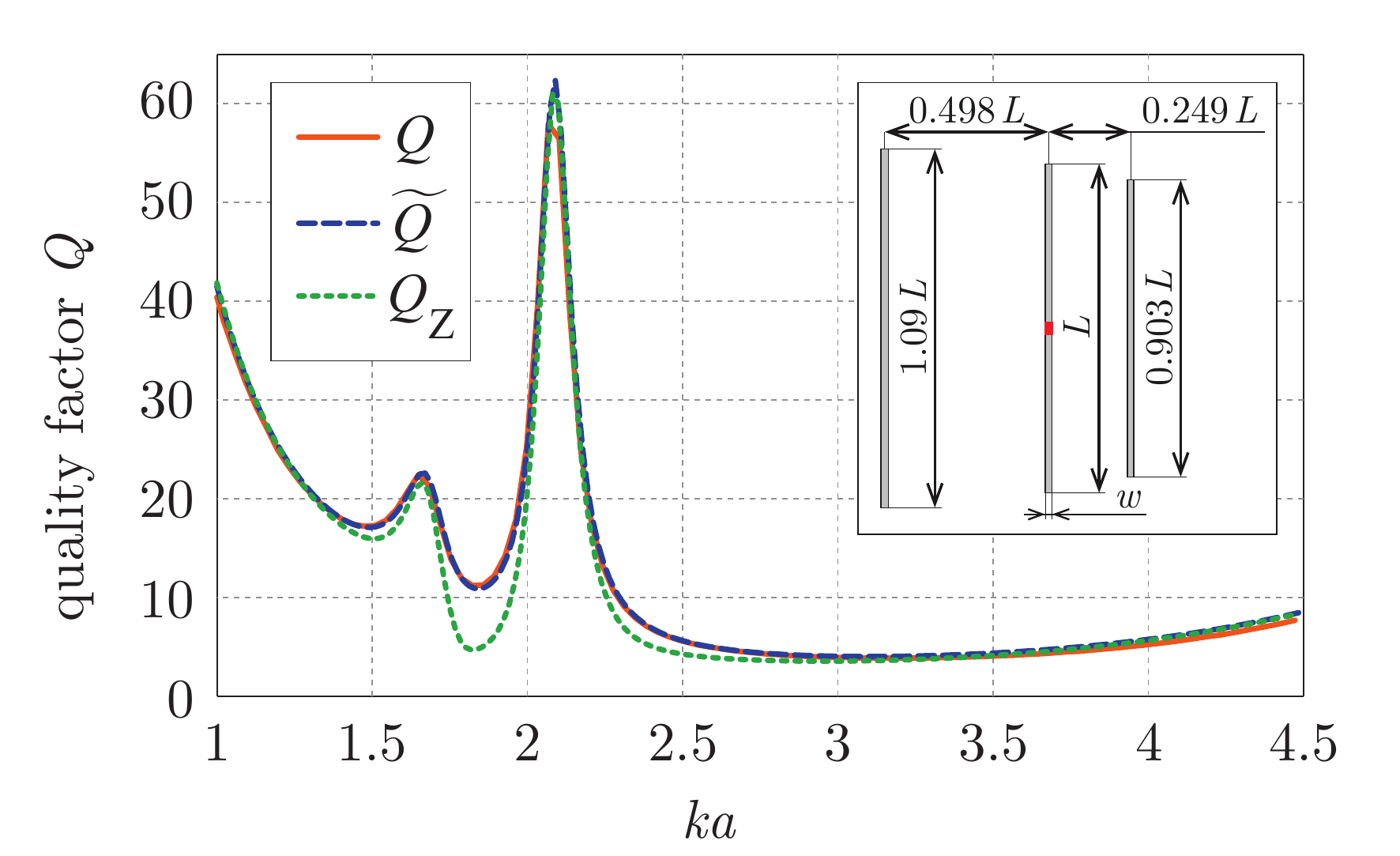}
\caption{Comparison of all Q factors of Yagi-Uda antenna (antenna proportions are stated in the inset).}
\label{fig:YagiFig2}
\EF

\section{Discussion}
\label{Sec_Disc}
Based on the previous sections, important properties of the novel time domain technique can be isolated and discussed. This discussion also poses new and so far unanswered questions that can be addressed in future.

The coordinate independence / dependence constitutes an important issue of many similar techniques evaluating the stored electromagnetic energy. Contrary to the radiation energy subtraction of Fante \cite{Fante_QFactorOfGeneralIdeaAntennas}, Rhodes \cite{Rhodes_AReactanceTheorem}, Yaghjian and Best \cite{YaghjianBest_ImpedanceBandwidthAndQOfAntennas}, or Gustafsson and Jonsson \cite{Gustaffson_StoredElectromagneticEnergy_PIER}, the new time-domain method can be proved to be coordinate-independent. It means that the same results are obtained irrespective of the position and rotation of the coordinate system. Due to the explicit reference to coordinates, this statement in question may not be completely obvious from (\ref{Basic_A08}). However it should be noted that any potential spatial shift or rotation of coordinate system emerges only as a static time shift of the received signal at the capturing sphere. Such static shift is irrelevant to the energy evaluation due to the integration over semi-infinite time interval.

The positive semi-definiteness represents another essential characteristic. It should be immanent in all theories concerning the stored energy. Although (\ref{Basic_A08}) contains the absolute value, it is difficult to mathematically prove the positive semi-definiteness of the stored energy evaluation as a whole, because it is not automatically granted that the integration during the second run integrates smaller amount of energy than the integration during the first run. Despite that, we can anticipate the expected behaviour from the physical interpretation of the method, which stipulates that the energy integrated in the second run must have been part of the first run as well. At worst, the subtraction of both runs can give null result. This observation is in perfect agreement with the numerical results. Nevertheless, the exact and rigorous proof admittedly remains an unresolved issue that is to be addressed in the future.

Unlike the methods of Fante \cite{Fante_QFactorOfGeneralIdeaAntennas}, Rhodes \cite{Rhodes_ObservableStoredEnergiesOfElectromagneticSystems} or Collin and Rothschild \cite{CollinRotchild_EvaluationOfAntennaQ}, the obvious benefit of the novel method consists in its ability to account for a shape of the radiator, not being restricted to the exterior of circumscribing sphere. 

Finally, it is crucial to realize that the novel method is not restricted to the time-harmonic domain, but can evaluate the stored energy in any general time-domain state of the system. This raises new possibilities for analyzing radiators in the time domain, namely the ultra-wideband radiators and other systems working in the pulse regime.

\section{Conclusion}
\label{Sec_Concl}
Three different concepts aiming to evaluate the stored electromagnetic energy and the resulting quality factor $Q$ of radiating system were investigated. The novel time domain scheme constitutes the first one, while the second one utilizes time-harmonic quantities and classical radiation energy extraction. The third one is based on the frequency variation of radiator?s input impedance. All methods were subject to in-depth theoretical comparison and their differences were presented on general non-radiating RLC networks as well as common radiators.

It was explicitly shown that the most practical scheme based on the frequency derivative of the input impedance generally fails to give the correct quality factor, but may serve as a very good estimate of it for structures that are well approximated by series or parallel resonant circuits. In contrast, the frequency domain concept with far-field energy extraction was found to work correctly in the case of general RLC circuits and simple radiators. Unlike the newly proposed time domain scheme, it could however yield negative values of stored energy, which is actually known to happen for specific current distributions. In this respect, the novel time domain method proposed in this paper could be denoted as reference, since it exhibits the coordinate independence, positive semi-definiteness, and most importantly, takes into account the actual radiator shape. Another virtue of the novel scheme is constituted by the possibility to use it out of the time-harmonic domain, e.g. in the realm of radiators excited by general pulse.

The follow-up work should focus on the radiation characteristics of separated parts of radiators or radiating arrays, the investigation of different time domain feeding pulses and their influence on performance of ultra-wideband radiators and, last but not least, on the theoretical formulation of the stored energy density generated by the new time domain method. 


\dataccess{This manuscript does not contain primary data and as a result has no supporting material associated with the results presented.}

\authorscontr{All authors contributed to the formulation, did numerical simulations and drafted the
	manuscript. All authors gave final approval for publication.}

\ack{The authors are grateful to Ricardo Marques (Department of Electronics and Electromagnetism, University of Seville) and Raul Berral (Department of Applied Physics, University of Seville) for many valuable discussions that stimulated some of the core ideas of this contribution. The authors are also grateful to Jan Eichler (Department of Electromagnetic Field, Czech Technical University in Prague) for his help with the simulations.}

\funding{The authors would like to acknowledge the support of COST IC1102 (VISTA) action and of project 15-10280Y funded by the Czech Science Foundation.}

\conflict{We declare to have no competing interests.}

\bibliographystyle{rspublicnat}
\bibliography{references_LIST}

\end{document}